\documentclass{aa}

\usepackage{graphicx}
\usepackage{txfonts}
\usepackage{xcolor}
\usepackage{hyperref}
\usepackage{amssymb}
\usepackage{units}
\usepackage{threeparttable}
\usepackage{natbib}
\usepackage{float}
\usepackage{adjustbox}

\begin{document}

   \title{ Kinematics and star formation of hub-filament systems in W49A}

   \author{WenJun Zhang
         \inst{1,2}  
          \and
   %%      %\fnmsep\thanks{Thanks.. if needed}
    Jianjun Zhou\inst{1,3,4} \and
    Jarken Esimbek \inst{1,3,4} \and
   Willem Baan\inst{1,5} \and
    Yuxin He\inst{1,3,4} \and
     Xindi Tang\inst{1,3,4} \and
     Dalei Li\inst{1,3,4} \and
     Weiguang Ji\inst{1,3,4} \and
     Gang Wu\inst{1,6} \and
     Yingxiu Ma\inst{1} \and
      Jiasheng Li\inst{1,2} \and
      Dongdong Zhou\inst{1} \and
      Kadirya Tursun\inst{1} \and
      Toktarkhan Komesh\inst{7,8}
      }

   \institute{XingJiang Astronomical Observatory, Chinese Academy of Sciences(CAS), Urumqi 830011, PR China \\ e-mail:
   zhangwenjun@xao.ac.cn; zhoujj@xao.ac.cn\
         \and
              University of the Chinese Academy of Sciences, Beijing,100080, PR China \ 
            \and Key Laboratory of Radio Astronomy, Chinese Academy of Sciences, Urumqi 830011, PR China\  
            \and Xinjiang Key Laboratory of Radio Astrophysics, Urumqi 830011, PR China  \
            \and Netherlands Institute for Radio Astronomy, ASTRON, 7991 PD Dwingeloo, The Netherlands \
            \and Max-Planck-Institut für Radioastronomie, Auf dem Hügel 69, D-53121 Bonn, Germany \
            \and Energetic Cosmos Laboratory, Nazarbayev University, Astana 010000, Kazakhstan \
            \and Faculty of Physics and Technology, Al-Farabi Kazakh National University, Almaty, 050040, Kazakhstan \
           }                            

   \date{Received day month year; Accepted ...}

% \abstract{}{}{}{}{} 
% 5 {} token are mandatory
\abstract 
{}
{W49A is a prominent giant molecular cloud (GMC) that exhibits strong star formation activities, yet its structural and kinematic properties remain uncertain. Our study aims to investigate the large-scale structure and kinematics of W49A, and elucidate the role of filaments and hub-filament systems (HFSs) in its star formation activity.}
{We utilized continuum data from \textit{Herschel} and the \textit{James Clerk Maxwell} Telescope (JCMT) as well as the molecular lines $^{12}$CO\,(3-2), $^{13}$CO\,(3-2), and C$^{18}$O\,(3-2) to identify filaments and HFS structures within W49A. Further analysis focused on the physical properties, kinematics, and mass transport within these structures. Additionally, recombination line emission from the H\,{\scriptsize I}/OH/Recombination (THOR) line survey was employed to trace the central H\,{\scriptsize II} region and ionized gas.}
{Our findings reveal that W49A comprises one blue-shifted (B-S) HFS and one red-shifted (R-S) HFS, each with multiple filaments and dense hubs. Notably, significant velocity gradients were detected along these filaments, indicative of material transport toward the hubs. High mass accretion rates along the filaments facilitate the formation of massive stars in the HFSs. Furthermore, the presence of V-shaped structures around clumps in position-velocity diagrams suggests ongoing gravitational collapse and local star formation within the filaments.}
{Our results indicate that W49A consists of one R-S HFS and one B-S HFS, and that the material transport from filaments to the hub promotes the formation of massive stars in the hub. These findings underscore the significance of HFSs in shaping the star formation history of W49A.}


   \maketitle

%-------------------------------------------------------------------
\section{Introduction}

\label{Intro}
Survey results from \textit{Herschel} have shown that filaments are ubiquitous in molecular clouds and that most dense clumps or cores are formed in filaments \citep{2010A&A...518L.102A,molinari2010clouds,li2016atlasgal,2018A&A...616A..78M,2020MNRAS.492.5420S} and play a key role in star formation \citep{palmeirim2013herschel}. 
Filaments can overlap further to form a hub-filament system \citep[HFS;][]{myers2009filamentary}. 
Recent case studies and statistical studies indicate that HFSs are the favorite sites of high-mass star formation \citep{trevino2019dynamics,zhou2022atoms,liu2023evidence,xu2023atoms,Ma2023A&A...676A..15M,yang2023direct,kumar2020unifying}. 
In gravitation-dominated and hierarchical collapsing molecular clouds, the material is transported through filaments toward the gravitational center, or hub. 
These hubs continue to accumulate mass from the surrounding filaments, making them optimal sites for the formation of high-mass stars or star clusters \citep{bonnell2003hierarchical,smith2009simultaneous,peretto2013global,vazquez2019global}. 

However, our understanding of the kinematics and dynamics of HFSs is still rather limited in terms of important aspects such as velocity gradients along filaments, the material transport from filaments to hubs, the role of dynamic filamentary networks in influencing star formation within clumps, and the impact of stellar feedback within the hubs on the HFSs \citep{trevino2019dynamics,wang2020formation,mookerjea2023spiral}. 
%Therefore, more studies on the HFSs are needed to enhance our understanding of high-mass star formation.

W49A is a giant molecular cloud (GMC) that comprises several active high-mass star-forming regions, including W49A-North (W49A-N), W49A-South (W49A-S), and W49A-Southwest (W49A-SW); it has a molecular gas mass of $\sim$\,2$\times10^{5}$\,M$_{\odot}$ (\citealt{urquhart2018atlasgal,galvan2013muscle}) and is at a distance of $\sim$\,11.1\,kpc \citep{zhang2013parallaxes}. There are many ultra-compact H\,{\scriptsize II} regions in W49A, all of which harbor high-mass zero-age main-sequence (ZAMS) stars \citep{de1997multifrequency}. The total stellar mass is $\sim$\,5\,-\,7$\times10^{4}\,$M$_{\odot}$ \citep{homeier2005massive}. 
W49A has two main velocity components, at $\sim$\,4 and $\sim$\,12\,km\,s$\rm ^{-1}$ (\citealt{mufson1977structure,miyawaki1986structure,miyawaki2009large,simon2001structure}). 
It has been suggested that the colocation of these two regions implies that they are moving toward each other -- either as two clouds colliding or as an inward-outward collapse of one cloud (\citealt{serabyn1993fragmentation,welch1987star}) -- or that they are the result of feedback from nearby H\,{\scriptsize II} regions \citep{peng2010w49a}. 
\citet{galvan2013muscle} suggested that the starburst of W49A most likely occurred because of a localized gravitational collapse of a HFS and that there are three such filament structures associated with W49A-N. 

Past studies have mostly focused on the correlation between the two cloud regions at $\sim$\,4 and $\sim$\,12\,km\,s$\rm ^{-1}$ and the cause of the active star formation \citep{peng2010w49a,galvan2013muscle,miyawaki2022star},  while paying little attention to their structure and kinematics. In this work we primarily investigate the structure, kinematics, and star formation of two major HFSs located on both sides of the central H\,{\scriptsize II} region in W49A. In Sect.\ref{Obs_and_lines} we introduce the data used in this study.  
In Sect.\ref{Results} we introduce the data processing results in detail. 
In Sect.\ref{Discussions} we discuss and analyze the observation results. 
Finally, in Sect.\ref{Conclusions} we summarize the main results of this work.

%--------------------------------------------------------------------

\begin{figure*}
    \centering
    \includegraphics[width=1.0\textwidth]{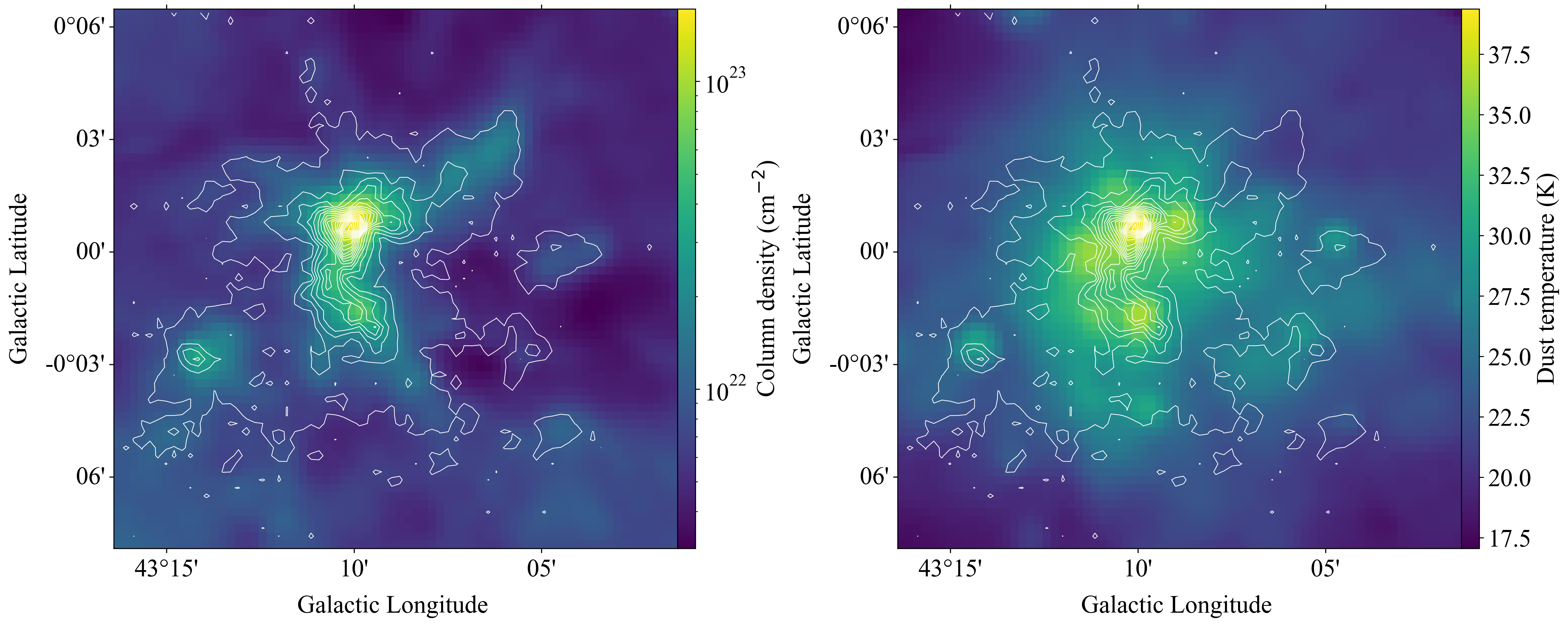}
    \caption{ Color map of the hydrogen molecular column density (left) and temperature (right) distributions of the W49A. The contours (from the outside in) represent 1.5\% to 96.5\% of the peak integrated intensity of $^{13}$CO\,(3-2), in intervals of 5\%.}
   \label{NH2 T_dust}
\end{figure*}

\section{Archive data}
\label{Obs_and_lines}
\subsection{The CO molecular data}
The $^{12}$CO\,(3-2) data were obtained from the CO High-Resolution Survey (COHRS) with an angular and spectral resolution of 14\,$\arcsec$ and 1\,km\,s$\rm ^{-1}$, respectively \citep{dempsey2013co}, and a root-mean-square (rms) sensitivity $\sigma$(T$_\text{A}$)\,$\approx$\,1\,K per channel \citep{rigby2019chimps}. 
The $^{13}$CO\,(3-2) and C$^{18}$O\,(3-2) data were obtained from the CO Heterodyne Inner Milky Way Plane Survey (CHIMPS), which has an angular and spectral resolution of 15\,$\arcsec$ and 0.5\,km\,s$\rm ^{-1}$\citep{rigby2016chimps}. 
The survey achieved mean rms sensitivities of $\sigma$(T$_\text{A}$)\,$\approx$\,0.6\,K and 0.7\,K per 0.5\,km\,s$\rm ^{-1}$ velocity channel for $^{13}$CO\,(3-2) and C$^{18}$O\,(3-2), respectively. 
These two surveys were performed with the \textit{James Clerk Maxwell} Telescope (JCMT) in Hawaii.

\subsection{The radio recombination line data}
The radio recombination lines (RRLs) H151$\alpha$\,-\,H156$\alpha$ and H158$\alpha$ (1.6\,-\,1.9\,GHz) were obtained from the H\,{\scriptsize I}/OH/Recombination (THOR) line survey of the inner Milky Way survey \citep{beuther2016hi,wang2020hi}. 
Observations of 5\,-\,6 minutes per pointing were conducted with the Very Large Array C configuration in L band \citep{beuther2016hi}. 
With a significant detection of all RRLs, the data were gridded to a spectral resolution of 5\,km\,s$\rm ^{-1}$ \citep{beuther2016hi}. The first release of the THOR data covered observations for $l$ = 14.0$^{\circ}$\,-\,37.9$^{\circ}$, and $l$ = 47.1$^{\circ}$\,-\,51.2$^{\circ}$, $\left| b \right| < 1.25^{\circ}$. The second release provided observations of the whole survey ($l$ = 14.0$^{\circ}$ - 67.4$^{\circ}$ and $\left| b \right| < 1.25^{\circ}$). Centre d'Analyse de Données Etendues (CADE)\footnote{\url{https://cade.irap.omp.eu/dokuwiki/doku.php?id=thor}} currently provides H\,{\scriptsize I} integrated intensity maps at a resolution of 40\,$\arcsec$ (excluding continuum) in units of Jy\,beam$\rm ^{-1}$\,km\,s$\rm ^{-1}$, and continuum emission maps at frequencies of 1060, 1310, 1440, 1690, 1820, and 1950\,MHz with a resolution of 25\,$\arcsec$, in units of Jy\,beam$\rm ^{-1}$ \citep{beuther2016hi,wang2020hi}. For the combined THOR+Very
Large Array Galactic Plane Survey (VGPS) data, the 1$\sigma$ brightness sensitivities for a spectral resolution of 1.6\,km\,s$\rm ^{-1}$ at 21\,$\arcsec$, 40\,$\arcsec$, and 60\,$\arcsec$ are 16, 3.9, and 1.8\,K, respectively. At 60\,$\arcsec$ resolution, the corresponding 1$\sigma$ rms of the VGPS data alone is even better than $\sim$\,1.5\,K \citep{beuther2016hi}. 

\subsection{The far-infrared data}
The \textit{Herschel} key project,  \textit{Herschel} infrared Galactic Plane Survey \citep[Hi-GAL;][]{molinari2010hi}, is the first unbiased survey of the galactic plane in the far infrared. 
Hi-GAL covers the entire Galactic plane with a nominal latitude limit of $\left| b \right| < 1^{\circ}$. 
The data include continuum images at 70, 160, 250, 350, and 500\,µm obtained with the Photodetector Array Camera and Spectrometer \citep[PACS;][]{poglitsch2010herschel} and Spectral and Photometric Imaging Receiver \citep[SPIRE;][]{griffin2010herschel} cameras on board the \textit{Herschel} Space Observatory \citep{pilbratt2010herschel}. 
The nominal beam sizes are 5.2\,$\arcsec$, 12\,$\arcsec$, 18\,$\arcsec$, 25\,$\arcsec$, and 37\,$\arcsec$ at 70, 160, 250, 350, and 500\,µm, respectively. 

\subsection{The mid-infrared data}
The Galactic Legacy Infrared Mid-Plane Survey Extraordinaire (GLIMPSE) is a mid-infrared survey (3.6, 4.5, 5.8, and 8.0\,µm) of the Inner Galaxy performed with the \textit{Spitzer} Space Telescope \citep{benjamin2003glimpse}.
The angular resolution is better than 2\,$\arcsec$ at all wavelengths \citep{benjamin2003glimpse}. 
The MIPS/\textit{Spitzer} Survey of the Galactic Plane (MIPSGAL) is a survey of the same region as GLIMPSE at 24 and 70\,µm, using the Multiband Imaging Photometer (MIPS) on board the \textit{Spitzer} Space Telescope \citep{rieke2004multiband}. 
The angular resolutions at 24 and 70\,µm are 6\,$\arcsec$ and 18\,$\arcsec$, respectively.

\section{Results}

\label{Results}

\subsection{The column density and dust temperature distribution}
\label{morphology}
We used \textit{Herschel} images to fit spectral energy distribution (SED) and obtain the target region's hydrogen molecule column density and dust temperature. Because some saturated pixels appear in the images at 160, 250, and 350\,µm in the center of W49A (see the top panel of Fig.\ref{fig:interpolating Herschel image}), it is necessary to recover the missed fluxes of those saturated pixels \citep{lin2016cloud}. Here we used the two-dimensional inward interpolation method on the original images of W49A at 160, 250, and 350\,µm to estimate the fluxes of those saturated pixels (see the bottom panel of Fig.\ref{fig:interpolating Herschel image}). 

The \textit{Herschel} data were employed to derive the temperature and column density map of W49A using pixel-by-pixel SED fitting \citep{wang2015large}. 
A Fourier transform was first performed on the original image to obtain high- and low-frequency components. 
The low-frequency components represented the background radiation and were subtracted from the data to remove the background from the image \citep{wang2015large}. 
Subsequently all images at 70, 160, 250, and 350\,µm were convolved to a circular Gaussian beam with full width at half maximum (FWHM) = 36.4\,$\arcsec$ using the kernels provided by \citet{aniano2011common} and re-gridded to the same pixel size. 
Finally, we fit each pixel with the following formula:
\begin{equation}
    I_{\nu} = B_{\nu} (1 - e^{-\tau_{\nu}}),
\end{equation}
where the Planck function, $B_{\nu}$, is modified by the optical depth \citep{kauffmann2008mambo}:
\begin{equation}
    \tau_{\nu} = \frac{\mu_{\text{H}_2} m_\text{H} \kappa_\nu N_{\text{H}_2}}{R_\text{gd}},  
\end{equation}
where $\mu_{\text{H}_2} = 2.33$ is the mean molecular weight adopted from \citet{kauffmann2008mambo}, $m_\text{H}$ is the mass of a neutron, $N_{\text{H}_2}$ is H$_2$ column density, $R_\text{gd}=100$ is gas to dust ratio, and the dust opacity per unit dust mass follows from \citep{ossenkopf1994dust}
\begin{equation}
    \kappa_\nu = 4.0 {(\frac{\nu}{505\text{GHz}})}^{\beta} \text{cm}^2\,\text{g}^{-1}, 
\end{equation}
where the dust emissivity index $\beta$ is fixed to 1.75 in the fitting \citep{wang2015large}. 
Finally, we obtained the column density and dust temperature maps, as shown in Fig.\ref{NH2 T_dust}. The hydrogen molecule column density and dust temperature calculated by us are consistent with that of \citet{lin2016cloud}. For the column density and dust temperature at the peak position ($l$ = 43.17$^{\circ}$, $b$ = 0.01$^{\circ}$), our fitting results are $1.7\times10^{23}$\,cm$^{-2}$ and 39.4\,K, respectively, while the corresponding values of \citet{lin2016cloud} are $2.4\times10^{23}$\,cm$^{-2}$ and 39.4\,K.

\subsection{Filaments and hub-filament systems}
\subsubsection{Optical depth of CO\,(3-2)}
We calculated the optical depth ($\tau$) of isotopic CO\,(3-2) lines in W49A. The optical depth of $^{12}$CO\,(3-2) is greater than 1 in most regions of W49A, while the optical depth of $^{13}$CO\,(3-2) is less than 1 in most areas (see Fig.\ref{tao_13CO}). Although C$^{18}$O\,(3-2) exhibits a significantly lower optical depth than 1, it only traces the densest region at the center of W49A. Therefore, we mainly use $^{13}$CO\,(3-2) for the analysis of this work.  

\subsubsection{Filaments}
\begin{figure}
    \centering
    \includegraphics[width=0.5\textwidth]{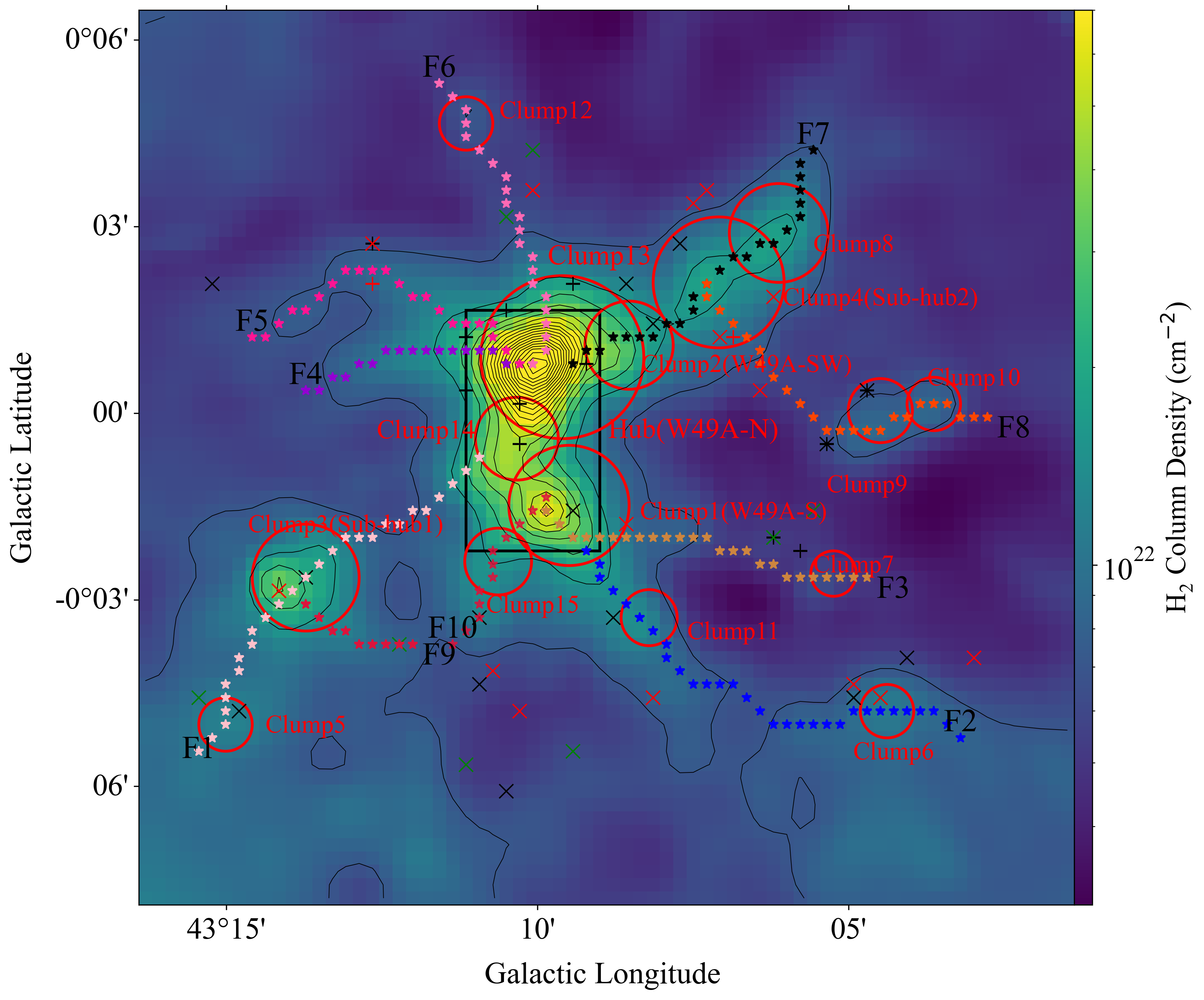}
    \caption{Filament structures of the W49A region. The color map is the distribution of the hydrogen molecular column density. The range of contours (from the outside in) is from $7.06\times10^{21}$\,cm$^{-2}$ to $1.71\times10^{23}$\,cm$^{-2}$. The star-like skeleton shows the filaments, and the circles and boxes represent W49A-N and the dense clumps from the catalog of  \citet{urquhart2014atlasgal} and \citet{de2021surveying}. Plus signs in the figure represent the YSOs marked by \citet{saral2015young} in W49, and the cross signs denote YSOs reported by \citet{kuhn2021spicy}. A black cross represents Class\,\uppercase\expandafter{\romannumeral1} YSO, a red cross is a Class\,\uppercase\expandafter{\romannumeral2} YSO, a green plus sign represents Class\,\uppercase\expandafter{\romannumeral3} YSOs, and a green cross sign is for a flat SED source.}
    \label{filament}
\end{figure}

Following \citet{wang2015large}, we identified filaments in W49A by considering the morphology, temperature, and velocity coherence (see Fig.\ref{CO channel map}). 
Filfinder\footnote{\url{https://fil-finder.readthedocs.io/en/latest/}} is a Python package integrated to identify filamentary structures in clouds \citep{2015MNRAS.452.3435K}. This method relies on a mask matrix. We used regions with $^{13}$CO\,(3-2) integrated intensity greater than five times the RMS as the mask matrix. Then, filamentary structures were identified as skeletons in the distribution of hydrogen molecule column density within the masked regions.
A total of ten filaments were found in W49A, and they are displayed in Fig.\ref{filament}. 

\subsubsection{The blue- and red-shifted hub-filament system}
\label{subsec:hfs}
The averaged $^{13}$CO\,(3-2) profiles of W49A (see Figs.\ref{dense spec} and \ref{dense core spec}) indicate that there are two main molecular clouds with different velocities along the line of sight: one blue-shifted (B-S) cloud (-4 to 7\,km\,s$\rm ^{-1}$) and one red-shifted (R-S) cloud (7 to 23\,km\,s$\rm ^{-1}$). These two clouds have been detected with many molecules at different optical depths, including $^{12}$CO, $^{13}$CO, C$^{18}$O, CS(2-1), H$^{13}$CO$^{+}$\,(1-0), H$^{13}$CN\,(1-0), and N$_2$H$^{+}$(1-0) \citep{mufson1977structure,miyawaki1986structure,galvan2013muscle}.
The system velocities of these two clouds are $\sim$\,4 and 12\,km\,s$\rm^{-1}$ , respectively \citep{miyawaki1986structure,serabyn1993fragmentation,miyawaki2022star}.

\begin{figure*}
    \centering
    \includegraphics[width=1.0\textwidth]{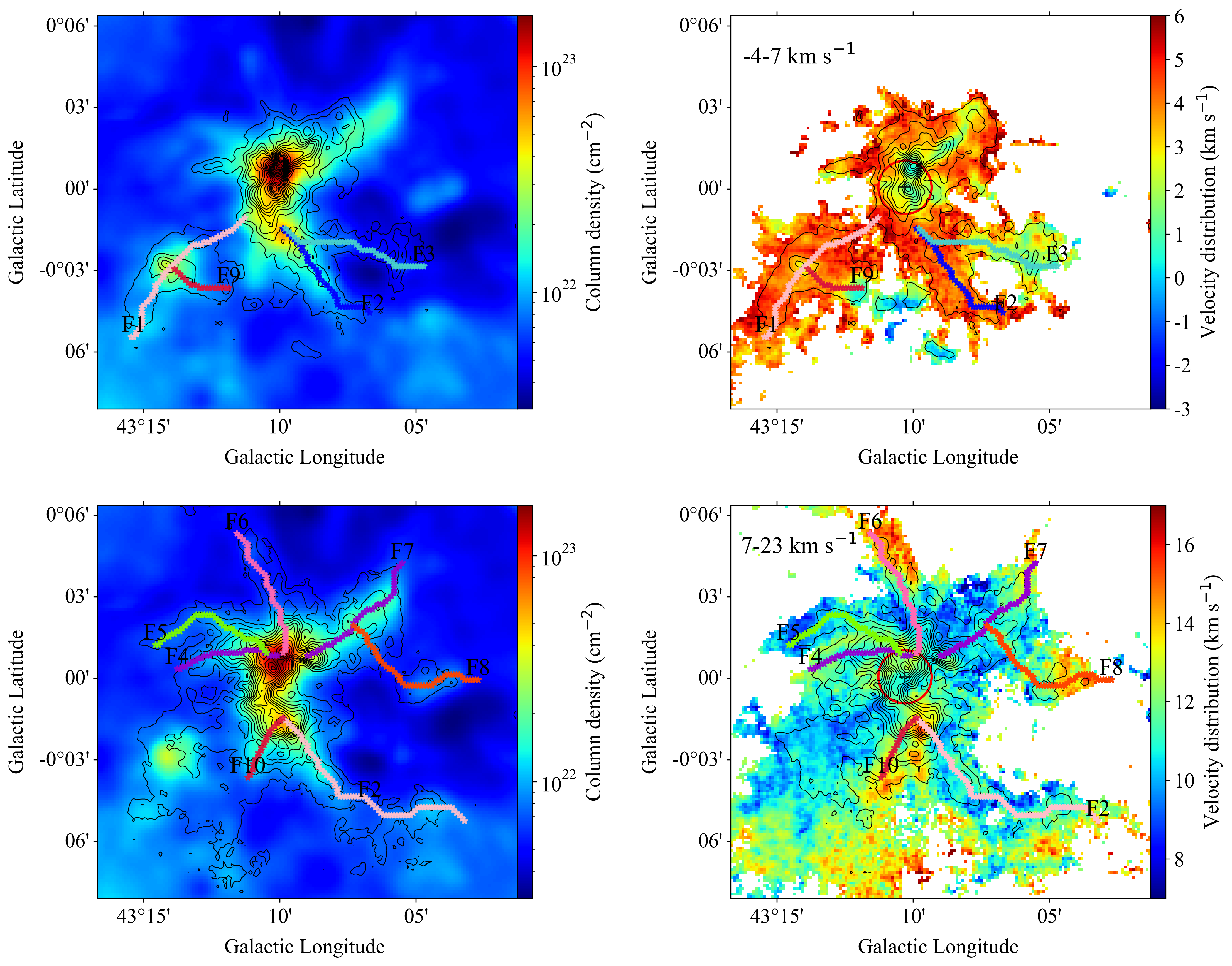}
    \caption{B-S and R-S HFSs in W49A. \textit{Left panels}: Heat map representing the hydrogen molecular column density overlaid with the integrated intensity contours of the R-S (bottom) and B-S (top) components of $^{12}$CO\,(3-2). The colored dots mark the skeletons of the filaments corresponding to the R-S (bottom) and B-S (top) components. 
    \textit{Right panels}: Heat map representing the velocity distribution of the R-S (bottom) and B-S (top) components of $^{12}$CO\,(3-2), overlaid with contours of the integrated intensity of the R-S (bottom) and B-S (top) components. The masked area is where the integrated intensity is less than ten times that of the image rms. The red circle indicates the shell of the H\,{\scriptsize II} region in W49A-N with the radius of $\sim$\,3.3\,pc, and the black cross indicates the central location \citep{peng2010w49a}.}
    \label{low(high) velo}
\end{figure*}

In the left panels of Fig.\ref{low(high) velo}, contours of the integrated intensity of $^{12}$CO\,(3-2) for B-S and R-S clouds are respectively superimposed on the hydrogen molecule column density map, with the corresponding filament skeletons marked. Filaments F\,1, F\,2, F\,3, and F\,9 match the B-S clouds well (upper-left panel of Fig.\ref{low(high) velo}). F\,2 and F\,3 converge toward the B-S dense clumps in W49A-S. F\,1 converges toward Clump\,14 in the hub, and F\,9 intersects with F\,1 at Clump\,3 (Sub-hub\,1). We refer to the HFS composed of F\,1, F\,9, and Clump\,3 as Sub-HFS\,1. All this suggests that the F\,1, F\,2, F\,3, F\,9, and B-S dense clumps in the hub constitute the B-S HFS.

Similarly, filaments F\,2, F\,4, F\,5, F\,6, F\,7, F\,8, and F\,10 match the R-S clouds (lower-left panel of Fig.\ref{low(high) velo}). Here F\,4, F\,5, and F\,6 converge to the dense clumps in W49A-N, F\,2 and F\,10 converge to W49A-S, F\,7 converges to W49A-SW, and F\,8 intersects with F\,7 at Clump\,4 (Sub-hub\,2). F\,7, F\,8, and Clump\,4 constitute Sub-HFS\,2. W49A-S and W49A-SW match the R-S cloud well and likely are part of it. We suggest that F\,2, F\,4, F\,5, F\,6, F\,7, F\,8, F\,10, and R-S dense clumps in W49A-N, W49A-SW, and W49A-S constitute the R-S HFS.

\begin{figure}
    \centering
    \includegraphics[width=0.5\textwidth]{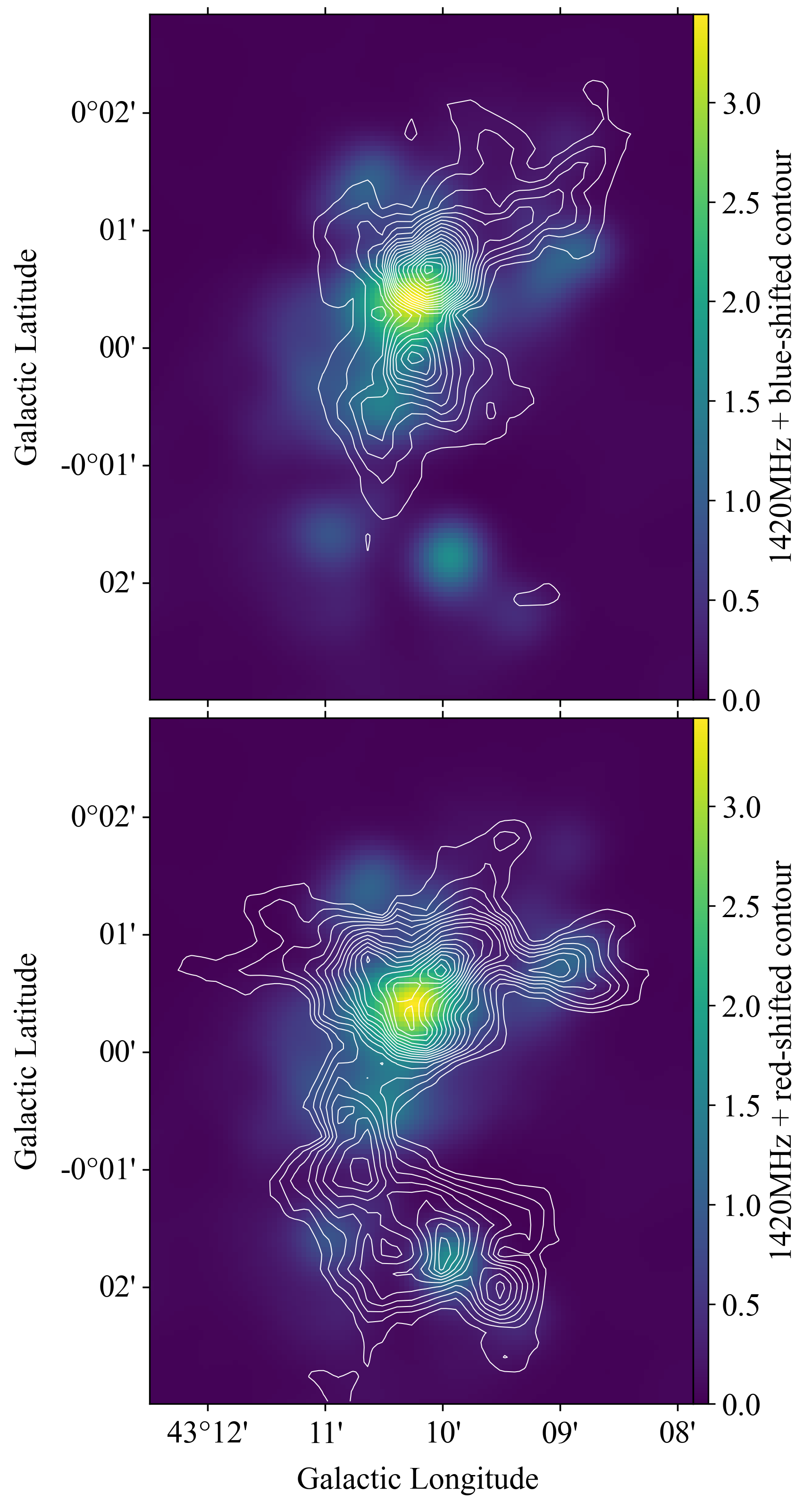}
    \caption{Integrated intensity of $^{13}$CO\,(3-2) for the B-S HFS (top) and R-S HFS (bottom), shown with the contour. The color map is the 1420\,MHz continuum emission.}
   \label{1420MHZ}
\end{figure}

\subsection{Properties of the filaments}
\label{filaments propertie}

\subsubsection{Velocity components and velocity dispersion in the filaments}
\label{filaments com}
The kinematics of the filaments within W49A, and spectra of $^{13}$CO\,(3-2) extracted at 27 different positions along the ten filaments are shown in Fig.\ref{filament_spec}. At positions outside the hub, spectra of $^{13}$CO\,(3-2) often show a single velocity component, while at positions near the hub, dual velocity components appear, such as at P\,4 on F\,2, P\,10 and P\,11 on F\,4, P\,12 on F\,5, and P\,18 on F\,7. Therefore, the B-S HFS coincides with the R-S HFS along the line of sight; they mainly appear in W49A-N, where their hubs coincide.

There are noticeable velocity gradients along the filaments. For instance, the peak velocities at P\,7, P\,8, and P\,9 of F\,3 are 7.44(±0.36), 2.33(±0.09), and 1.63(±0.14)\,km\,s$\rm ^{-1}$, respectively.  
Similarly, the peak velocities at positions P\,15, P\,16, and P\,17 of F\,6 are 8.94(±0.21), 13.10(±0.16), and 14.84(±0.47)\,km\,s$\rm ^{-1}$, respectively (see Fig.\ref{filament_spec}). All velocity gradient along the filaments and toward the hubs are presented in Table\,\ref{tab:filaments_properties}. 

\begin{table*}[!t]
\centering
\renewcommand\arraystretch{1.2}
\caption{Filaments and clumps in W49A.}
\label{tab:filaments_properties}
\resizebox{\linewidth}{!}{
\begin{tabular}{ccccc}
\hline
\hline
Filament ID & Affiliated HFS & Structures associated with the filament & Velocity gradients along the filament & Velocity gradient near the clump$\tnote{a}$ \\ 
          &   & (clumps and filaments)  & (km\,s$\rm ^{-1}$\,pc$\rm ^{-1}$)  & (km\,s$\rm ^{-1}$\,pc$\rm ^{-1}$)\\
\hline
F\,1 & B-S HFS, Sub-HFS\,1 & C\,3, C\,5, F\,9 & -0.42(0-6.8\,pc), -0.09(17.3-22.3\,pc) & -1.48 and 1.08(C\,3) \\

F\,2(B-S) & B-S HFS & W49A-S & 0.3(0-10\,pc),-0.67(10-13\,pc) & - \\

F\,2(R-S) & R-S HFS & W49A-S, C\,11, C\,6 & 0.29(20-28\,pc) & -2.37 and 0.58(W49A-S),-0.95 and 0.24(C11) \\

F\,3 & B-S HFS & W49A-S, C\,7 & -2.13(0-4.3\,pc), -1.96(4.3-7.4\,pc) &  -0.20 and 0.31(C\,7) \\

F\,4 & R-S HFS & W49A-N & 1.25(0-3.7\,pc), 0.22(3.7-13\,pc) & - \\

F\,5 & R-S HFS & W49A-N & 0.92(0-2.8\,pc),-0.27(2.8-18\,pc) & - \\

F\,6 & R-S HFS & W49A-N, C\,12 & 0.39(0-17.5\,pc) & -0.28 and 0.25(C\,12) \\

F\,7 & R-S HFS, Sub-HFS\,2 & W49A-SW, C\,8, C\,4, F\,8 & 0.31(5.5-21\,pc) & -0.88 and 1.21(W49A-SW),-0.41 and 0.7(C4) \\

F\,8 & R-S HFS, Sub-HFS\,2 & C\,4, C\,9, C\,10, F\,7 & 0.27(5.2-17.5\,pc) & - \\

F\,9 & B-S HFS, Sub-HFS\,1 & C\,3, F\,1 & -0.11(0-5\,pc),-2.72(5.5-7.2\,pc) & - \\

F\,10 & R-S HFS & W49A-S, C\,15 & -1.32(0-3.1\,pc),0.38(3.5-10\,pc) & - \\

\hline
\end{tabular}}
\end{table*}

The velocity dispersion (FWHM/$\sqrt{8\ln2}$) across the 27 positions on the filaments ranges from 0.83(±0.11) to 4.11(±0.21)\,km\,s$\rm ^{-1}$. 
These values are relatively large compared with that detected in other HFSs, for example 0.24\,-\,0.39\,km\,s$\rm ^{-1}$ for Monoceros R2 \citep{trevino2019dynamics},  0.58\,-\,1.49\,km\,s$\rm ^{-1}$ for G18.88-0.49 \citep{dewangan2020new}, and 0.58\,km\,s$\rm ^{-1}$ for G326.611+0.811 \citep{he2023investigating}. 
This increased dispersion may in part be due to the dual velocity gas components found and to the gas flow along the filaments. 
Indeed the velocity dispersion is larger when closer to a hub (for instance, on F\,1 in B-S HFS, the velocity dispersions for P\,1, P\,2, and P\,3 are 2.41(±0.23), 2.21(±0.19), and 1.1(±0.23)\,km\,s$\rm ^{-1}$, respectively. On F\,10 in R-S HFS, the velocity dispersions for P\,26 and P\,27 are 2.08(±0.16) and 1.66(±0.10)\,km\,s$\rm ^{-1}$, respectively.)

\subsubsection{The physical properties of the filaments}
\label{filaments physical properties}
Because there is mutual contamination caused by the overlap of the B-S and R-S HFSs in the line of sight in the hub regions and sections of F\,2 and F\,4 (Sect.\ref{filaments com}), we primarily calculated the physical properties of filaments F\,1, F\,3, F\,5, F\,6, F\,7, F\,8, F\,9, and F\,10, which are outside the B-S and R-S hub regions. It is necessary to clarify here that, due to the superposition of double velocity components in the line of sight near W49A-N (hub), the calculated lengths, masses, temperatures, and mass accretion rate of filaments are all values excluding those from the central hub. This may represent a lower limit of the actual values.

\begin{table}
\centering
\caption{Physical parameters of the filaments.}
\label{tab:physical_parameter_filament}
\renewcommand\arraystretch{1.2}
\begin{adjustbox}{width=0.5\textwidth,center}
\begin{tabular}{c c c c c} % four columns, alignment for each
\hline
\hline
Source ID & $l$ & $M$ & $T_\mathrm{dust}$  & $\dot{M}_\text{acc}$\\ 
& (pc) & ($\text{M}_\odot$) & (K) & ($\text{M}_\odot$\,year$^{-1}$)\\
\hline
F\,1 & 23.82$\pm1.55$ & 36027$\pm4682$ & 26.46 & -7.66$(\pm0.99)\times10^{-3}$ \\
F\,3 & 12.43$\pm0.81$ & 4237$\pm551$ & 26.05 & -8.74$(\pm1.14)\times10^{-4}$ \\
F\,5 & 15.66$\pm1.02$ & 11197$\pm1455$ & 25.80 & -3.24$(\pm0.42)\times10^{-3}$ \\
F\,6 & 13.92$\pm0.90$ & 7638$\pm993$ & 25.68 & 3.41$(\pm0.44)\times10^{-3}$ \\
F\,7 & 17.10$\pm1.11$ & 28092$\pm3651$ & 24.71 & 8.17$(\pm1.06)\times10^{-3}$ \\
F\,8 & 18.72$\pm1.22$ & 9562$\pm1243$ & 25.14 & 1.13$(\pm0.15)\times10^{-3}$ \\
F\,9 & 6.66$\pm0.43$ & 5681$\pm738$ & 24.66 & -3.26$(\pm0.42)\times10^{-3}$ \\
F\,10 & 6.69$\pm0.45$ & 6577$\pm855$ & 30.13 & 2.72$(\pm0.35)\times10^{-3}$ \\
\hline
\end{tabular}
\end{adjustbox}
\end{table}

Table\,\ref{tab:physical_parameter_filament} presents the physical properties of several filaments, including the length, mass, average temperature, and mass accretion rate on the filaments. 
The mass of the filaments was calculated based on the column density using the method proposed by \citet{Ma2023A&A...676A..15M}:
\begin{equation}
    M = \mu m_\text{H} \sum_{i}A_\text{pixel}(i) N_{\text{H}_{2}}(i),
    \label{eq:file_mass}
\end{equation}
where $A_\text{pixel}(i)$ represents the area of a pixel, and $N_{\text{H}_{2}}(i)$ is the hydrogen molecule column density corresponding to the pixel. 
According to Table\,\ref{tab:physical_parameter_filament}, the length range of the filament skeletons is estimated as 6.66($\pm0.43$)\,-\,23.82($\pm1.55$)\,pc, and the mass range is 4237($\pm551$)\,-\,36027($\pm4682$)\,$\text{M}_\odot$. 
The filaments in W49A have relatively large linear scales, contain significantly more mass, and provide an excellent environment for subsequent star formation. 

The formalism proposed by \citet{kirk2013filamentary} was used to estimate the accretion rate of filaments in W49A using a cylindrical model and velocity gradients as follows:
\begin{equation}
    \dot{M} = (\nabla V M) / \tan{\beta},
\label{macc}
\end{equation}
where $M$ represents the mass of the filament, and $\beta$ is the inclination angle of the filament on the plane of the sky, typically assumed to be 45\,$^{\circ}$. Using this method, we obtained the mass accretion rates for these filaments from the molecular cloud edge to the outer regions of the hub, with absolute values ranging from 8.74$(\pm1.14)\times$10$^{-4}$ to 8.17$(\pm1.06)\times$10$^{-3}$\,$\text{M}_\odot$\,year$^{-1}$ (see Table\,\ref{tab:physical_parameter_filament}). The filaments in W49A exhibit higher mass accretion rates compared with many other HFSs in the literature, for example 0.30\,-\,1.80$\times$10$^{-4}$\,$\text{M}_\odot$\,year$^{-1}$ for Monoceros\,R2 \citep{trevino2019dynamics}, 1.4\,-\,3.6$\times$10$^{-4}$\,\text{M}$_\odot$\,year$^{-1}$ for G310.142+0.758 \citep{yang2023direct}, and 1.2\,-\,3.6$\times$10$^{-4}$\,$\text{M}_\odot$\,year$^{-1}$ for G326.27-0.49 \citep{mookerjea2023spiral}. This is consistent with the fact that W49A is one of the strongest star-forming regions in the Milky Way.

\subsection{Star formation in the blue- and red-shifted hub-filament systems}
\label{dense structure}

In Fig.\ref{1420MHZ} the integrated $^{13}$CO\,(3-2) emission of both the B-S and R-S HFSs is overlaid on the color map of the 1420\,MHz continuum emission, respectively. The map shows that the 1420\,MHz continuum emission is a better match to the high-mass star-forming regions associated with the R-S HFS, such as W49A-N, W49A-SW, and W49A-S, but does not match the dense clumps of the B-S HFS; the strongest 1420\,MHz emission is between two dense clumps. However, compared with the observations with high sensitivity and high angular resolution at 3.6\,cm \citep{de1997multifrequency}, we find that most dense clumps in the R-S and B-S HFSs contain ultra-compact H\,{\scriptsize II} regions. There are many high-mass ZAMS O-type stars in these ultra-compact H\,{\scriptsize II} regions \citep{de1997multifrequency}. 
The distribution of the H$_\alpha$(n=151\,-\,158) emission, which traces the ionizing gas of the H\,{\scriptsize II} region, also matches the 1420\,MHz continuum emission (see Fig.\ref{rrl}) and shows the presence of an expanding H\,{\scriptsize II} region. 

\begin{figure}
    \centering
    \includegraphics[width=0.5\textwidth]{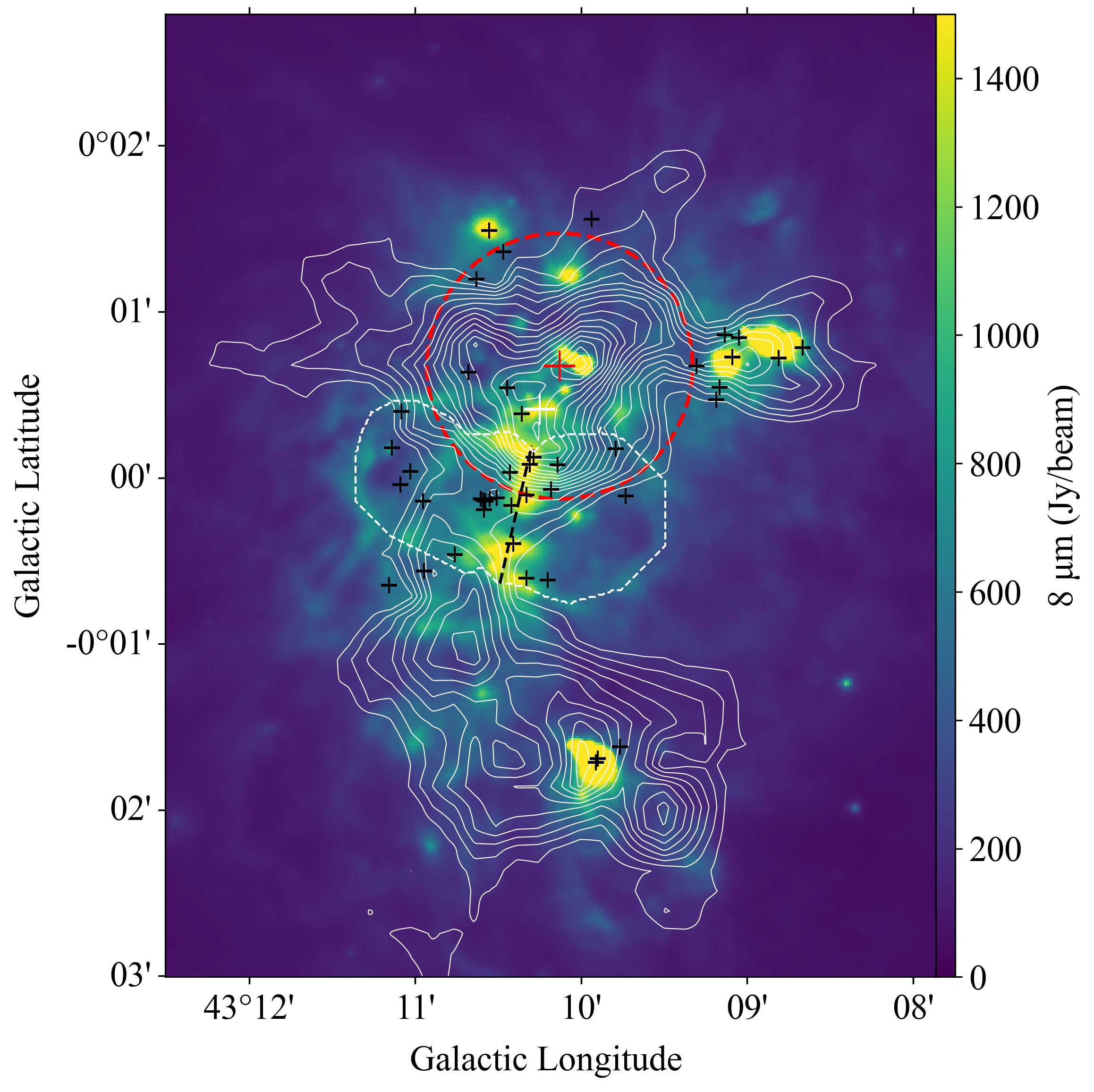}
    \caption{Color map 8\,µm continuum superposed on the white contours of the $^{13}$CO\,(3-2) integrated intensity of the R-S HFS. The red circles mark the collision of two dense components, as identified by \citet{miyawaki2022star}. The black plus signs indicate the distribution of massive stars in W49 confirmed by \citet{wu2016massive}. The dotted black line denotes the waist of the bipolar bubble, and the white plus sign is where the 1420\,MHz continuum radiation is strongest. The dashed white lines mark the outline of the bipolar bubble.}
    \label{8 dense}
\end{figure}

As shown in Fig.\ref{8 dense}, the distribution of the 8\,µm emission in W49A is also consistent with the R-S HFS. The strongest 1420\,MHz continuum emission is at the position of the densest clump of the R-S HFS and coincides with the waist of a bipolar bubble structure traced by 8\,µm emission. The dense gas of the R-S HFS overlaps with this waist region where most massive stars are located generating the bipolar bubble. 

%-----------------------------------------------------------------
\section{Discussion}
\label{Discussions}

\begin{table*}
\centering
\renewcommand\arraystretch{1.2}
%\begin{threeparttable}
\caption{Dense structures and their physical parameters in W49A.}
\label{tab:dense_structures}
\begin{tabular}{lccccccccc}
\hline
\hline
$\mathrm{Name}$ & $l_\text{cen} \tablefootmark{a}$ & $b_\text{cen} \tablefootmark{a} $ & ${R_\text{eq} \tablefootmark{b}}$ & $M \tablefootmark{c}$ & $T_\text{dust} \tablefootmark{d}$& $\mathrm{\Sigma \tablefootmark{e} } $ & ${n}(\text{H}_{2}) \tablefootmark{f}$ \\
       & ($^{\circ}$) & ($^{\circ}$) & (pc) & ($\text{M}_\odot$)& (K) & (g\,cm$\rm ^{-2}$) &  ($\times10^{3}$cm$ ^{-3}$)\\
\hline
W49A-N   & 43.17 & -0.01 & 5.46$\pm0.35$	& 98844$\pm12848$ & 33.86 & 0.22$\pm0.04$ & 2.10$\pm0.27$ \\
W49A-S & 43.16 & -0.03 & 3.66$\pm0.24$ & 26240$\pm3411$ & 31.91 & 0.13$\pm0.02$ & 1.85$\pm0.24$ \\
W49A-SW & 43.15 & 0.01 & 2.85$\pm0.19$ & 13553$\pm1762$ & 31.84 & 0.11$\pm0.02$ & 2.03$\pm0.26$ \\
Sub\_hub1 & 43.23 & -0.05 & 3.25$\pm0.21$ & 12745$\pm1657$ & 25.39 & 0.08$\pm0.01$ & 1.28$\pm0.17$ \\
Sub\_hub2 & 43.12 & 0.03 & 4.07$\pm0.26$ & 14858$\pm1931$ & 24.89 & 0.06$\pm0.01$ & 0.76$\pm0.09$ \\
Clump5 & 43.25 & -0.09 & 1.63$\pm0.11$ & 2090$\pm272$ & 20.74 & 0.05$\pm0.01$ & 1.67$\pm0.22$ \\
Clump6 & 43.08 & -0.08 & 1.63$\pm0.11$ & 1913$\pm249$ & 21.69 & 0.05$\pm0.01$ & 1.53$\pm0.20$ \\
Clump7 & 43.09 & -0.05 & 1.22$\pm0.08$ & 610$\pm79$ & 25.92 & 0.03$\pm0.01$ & 1.16$\pm0.15$ \\
Clump8 & 43.11 & 0.04 & 3.04$\pm0.20$ & 8385$\pm1089$ & 21.73 & 0.06$\pm0.01$ & 1.03$\pm0.13$  \\
Clump9 & 43.08 & 0.00 & 1.82$\pm0.12$ & 1823$\pm237$ & 26.48 & 0.04$\pm0.01$ & 1.05$\pm0.14$ \\
Clump10 & 43.06 & 0.00 & 1.63$\pm0.11$ & 1455$\pm189$ & 23.51 & 0.04$\pm0.01$ & 1.17$\pm0.15$  \\
Clump11 & 43.14 & -0.06 & 1.63$\pm0.11$ & 2173$\pm282$ & 24.62 & 0.05$\pm0.01$ & 1.74$\pm0.23$ &  \\
Clump12 & 43.19 & 0.07 & 1.63$\pm0.11$ & 1119$\pm145$ & 23.39 & 0.03$\pm0.01$ & 0.89$\pm 0.12$  \\
Clump13 & 43.16 & 0.01 & 5.08$\pm0.33$ & 83714$\pm10881$ & 33.57 & 0.22$\pm0.04$ & 2.20$\pm 0.29$  \\
Clump14 & 43.18 & -0.01 & 2.64$\pm0.17$ & 21777$\pm2831$ & 33.84 & 0.21$\pm0.04$ & 4.10$\pm 0.53$  \\
Clump15 & 43.18 & -0.04 & 2.03$\pm0.13$ & 4389$\pm571$ & 31.01 & 0.07$\pm0.01$ & 1.80$\pm 0.23$  \\
\hline
\end{tabular}
%\begin{tablenotes}
\tablefoot{
\tablefoottext{a}{Coordinates the dense structure's central point (Galactic longitude, latitude).} 
\tablefoottext{b}{The equivalent radius of the dense structure, $ R_\text{eq} = ({A_\text{clump}}/{\pi}) ^ {0.5} $\citep{rigby2019chimps}.} 
\tablefoottext{c}{The mass of dense structures is calculated from the hydrogen molecular column density.}
\tablefoottext{d}{The average dust temperature of dense structures.}
\tablefoottext{e}{The surface density of the dense structures calculated by $\Sigma = {M}/(\pi R_\text{eq}^{2})$ \citep{mazumdar2021high}.}
\tablefoottext{f}{The hydrogen molecular number density of the dense structure, calculated as $ \overline{n}(\text{H}_{2}) = {3}{M}/({4\pi}{\mu}m_\text{p}{R_\text{eq}}^{3})$ \citep{rigby2019chimps}.}
%\end{tablenotes}
%\end{threeparttable}
}
\end{table*}

\subsection{Kinematics and star formation in the blue-shifted hub-filament system}
\label{low-velocity HFS}
The center of the B-S HFS region has an interesting centripetal velocity gradient (see Fig.\ref{low high v}). The velocity decreases from $\sim$\,5.0 to 1.5\,km\,s$\rm ^{-1}$ from the edge of W49A-N to the dense clumps in the center of it, which suggests that the dense clumps of the B-S HFS are gravitationally collapsing. Figure\,\ref{lc1_lc2} illustrates the gas velocity structure around the two dense clumps, LC\,1 and LC\,2, in the B-S hub. The fitting results suggest that the gas appears to be converging toward the gravitational center under the influence of gravity \citep{hacar2017gravitational}, which further supports our idea. The strong 3.6\,cm radio emission within the LC\,1 and LC\,2 regions and many high-mass ZAMS stars identified in them \citep{de1997multifrequency} indicate that they are ultra-compact H\,{\scriptsize II} regions, while gas accretion at larger scales continues. 

Filament F\,1 show velocity gradients $\sim$\,-0.21\,km\,s$\rm ^{-1}$\,pc$\rm ^{-1}$ from the ends to the edge of Clump\,14 (hub in W49A-N; see Fig.\ref{f1_f9_pv}). F\,2 and F\,3 converge toward the B-S dense clumps in W49A-S, and show velocity gradients $\sim$\,0.30 and -0.17\,km\,s$\rm ^{-1}$\,pc$\rm ^{-1}$ from their ends to the edge of W49A-S, respectively (see Figs.\ref{f2_pv} and \ref{f3_pv}).
Both filaments probably transfer materials onto the B-S HFS dense clumps in the hub. Still, because of confusion with the R-S HFS along the line of sight, we only calculated the mass accretion rates from the periphery to the hub edge in F\,1 and F\,3, which are -7.66$(\pm0.99)\times10^{-3}$ and -8.74$(\pm1.14)\times10^{-4}$\,\text{M}$_\odot$\,year$^{-1}$, respectively (see Table\,\ref{tab:physical_parameter_filament}). This value is much higher than those derived for several other HFSs (see Sect.\ref{filaments physical properties}).  
Considering that F\,2 may also transfer materials toward the hub of the B-S HFS at a similar rate, the true mass accretion rate toward the hub may be $\sim$\,1.3$\times10^{-2}$\,\text{M}$_\odot$\,year$^{-1}$. This would indicate a significant inflow of material from the surrounding regions into W49A-N and W49A-S.

On the other hand, as anticipated by some numerical simulation models in the past \citep{bonnell2003hierarchical,smith2009simultaneous,vazquez2019global}, F\,1, F\,2, F\,3, and F\,9 may also fragment and form dense clumps locally. 
Such dense clumps may also accrete materials from the filaments and form low- and intermediate-mass stars. 
To check dense clumps in the filaments and their star formation, we extracted 15 clumps associated with the B-S and R-S HFSs from the catalog of \citet{urquhart2014atlasgal}. 
Table\,\ref{tab:dense_structures} shows their location, size, mass, and other physical properties. 
Because $^{13}$CO\,(3-2) may be optically thick in dense clumps, we did not try to derive their velocity dispersion and virial parameters. In fact, we checked the mass-size relationship ($M(r) \leq 580$ $\text{M}_\odot$ $(R_\text{eq}$ pc$^{-1})^{1.33}$ \citep{kauffmann2010many}) of these clumps and found that 14 of them satisfy the condition to form high-mass stars or star clusters.

Clump\,3 and filaments F\,1 and F\,9 constitute Sub-HFS\,1 (Sect.\ref{subsec:hfs}), and F\,1 shows a typical V-shaped structure around it in a position-velocity (P-V) diagram (see Fig.\ref{f1_f9_pv}). 
The two sides have velocity gradients of $\sim$\,-1.48 and 1.08\,km\,s$\rm ^{-1}$\,pc$\rm ^{-1}$, respectively. 
Such a V-shaped structure indicates that gravitational collapse is taking place there \citep{hacar2017gravitational,Ma2023A&A...676A..15M}. 
Similarly, V-shaped structures are also found around Clump\,7 in the P-V diagram of F\,3 (see Fig.\ref{f3_pv}). The velocity gradients along the two sides of Clump\,7 is $\sim$\,-0.20 and 0.31\,km\,s$\rm ^{-1}$\,pc$\rm ^{-1}$. All such clumps appear to be accreting materials from the filaments and are forming stars.

\subsection{Kinematics and star formation in red-shifted hub-filament system}

The R-S HFS also shows a centripetal velocity gradient in its hub (W49A-N) with a velocity increasing from $\sim$\,10 to 12\,km\,s$\rm ^{-1}$ from the edge of W49A-N to its center (see Fig.\ref{low high v}). Similarly, within the dense clumps of the R-S hub, there is also a significant presence of high-mass ZAMS stars \citep{de1997multifrequency}, many of which are associated with ultra-compact H\,{\scriptsize II} regions. However, compared to the B-S hub, the centripetal velocity gradient of the clumps in the R-S hub is less pronounced, and gas accretion around the clumps appears to be less evident. The filaments F\,4, F\,5, and F\,6 show velocity gradients $\sim$\,0.22, -0.27, and 0.39\,km\,s$\rm ^{-1}$\,pc$\rm ^{-1}$ from their ends to the edge of W49A-N, respectively (see Figs.\ref{f4_f5_pv} and \ref{f6_f7_pv}). They probably transfer materials onto the R-S dense clumps in W49A-N. Filament F\,7 also shows a velocity gradient $\sim$\,0.31\,km\,s$\rm ^{-1}$\,pc$\rm ^{-1}$ from its end to the edge of W49A-SW (see Fig.\ref{f6_f7_pv}). The velocity gradient along F\,8 from its end to the edge of Clump\,4 is $\sim$\,0.27\,km\,s$\rm ^{-1}$\,pc$\rm ^{-1}$ (Sub-hub\,2; see Fig.\ref{f8_f10_pv}). F\,10 shows a velocity gradient $\sim$\,0.38\,km\,s$\rm ^{-1}$\,pc$\rm ^{-1}$ from its end to the edge of W49A-S (see Fig.\ref{f8_f10_pv}). There is no clear velocity gradient along F\,2 from its end to the edge of W49A-S (see Fig.\ref{f2_pv}), which may be due to the projection effect. These velocity gradients along the filaments suggest the material transport onto the dense clumps in W49A-N, W49A-S, and W49A-SW.

The mass accretion rates for filaments F\,4, F\,5, and F\,6 from its end to the edge of W49A-N are $\sim$\,1.25$(\pm0.16)\times10^{-3}$, -3.24$(\pm0.42)\times10^{-3}$ and 3.41$(\pm0.44)\times10^{-3}$\,\text{M}$_\odot$\,year$^{-1}$, respectively. The mass accretion rates for F\,7 from its end to the edge of W49A-SW is $\sim$\,8.17$(\pm1.06)\times10^{-3}$\,\text{M}$_\odot$\,year$^{-1}$. The material transport rate from filament F\,10 to W49A-S is $\sim$\,2.72$(\pm0.35)\times10^{-3}$\,\text{M}$_\odot$\,year$^{-1}$. Such values are also much higher than the mass accretion rates found in other HFSs (10$^{-4}$\,-\,10$^{-3}$\,\text{M}$_\odot$\,year$^{-1}$; \citealt{2014prpl.conf...27A,trevino2019dynamics,yang2023direct,mookerjea2023spiral,Seshadri2024MNRAS.527.4244S}). Therefore, filaments of the R-S HFS also transfer materials from the surrounding regions onto the dense clumps in the center at a high rate.  

Similar to the B-S HFS, many dense clumps on the filaments of the R-S HFS show evidence of accreting materials from filaments locally. Typical V-shaped structures were detected around Clump\,1 (W49A-S) and Clump\,11 on F\,2, Clump\,12 on F\,6, Clump\,2 (W49A-SW) and Clump\,4 on F\,7, the velocity gradients of their two sides are  $\sim$\,-2.37 and 0.58, $\sim$\,-0.95 and 0.24, $\sim$\,-0.28 and 0.25, $\sim$\,1.21 and -0.88, $\sim$\,0.70 and -0.41\,km\,s$\rm ^{-1}$\,pc$\rm ^{-1}$, respectively (see Figs.\ref{f2_pv} and \ref{f6_f7_pv}). 
All these R-S HFS clumps are also accreting materials from the filaments and are forming stars.

Above results indicate that filaments not only transfer material into hub of the HFS, but also fragment and form dense clumps and further stars locally. In addition, most Class\,\uppercase\expandafter{\romannumeral1} young stellar objects (YSOs) are distributed on the hub or the dense clumps along the filament, while most Class\,\uppercase\expandafter{\romannumeral2} YSOs are distributed near the filaments (see Fig.\ref{filament}). These characteristics are consistent with the global hierarchical collapse model \citep{vazquez2019global}.

\subsection{Gravitational collapse and accretion flows in Sub-HFS\,1}
\label{Sub-HFS1}
In Sect.\ref{subsec:hfs} two local hub systems, Sub-HFS\,1 and Sub-HFS\,2, have been identified within the B-S and R-S HFSs, respectively. 
Because Sub-HFS\,1 is relatively isolated and is less affected by the complex environment, we focused on analyzing the gravitational collapse and accretion flow in Sub-HFS\,1. 

\subsubsection{A global velocity gradient in Sub-HFS\,1}
The kinematic characteristics along the filaments of Sub-HFS\,1 were obtained from Gaussian fitting of the molecular spectral lines of $^{13}$CO\,(3-2) along the filament skeleton. 
This resulted in the velocity distribution along the filaments F\,1 and F\,9 as shown in Fig.\ref{f1_f9_pv}.
The velocity variation along F\,1 is similar to that found in \citet{Ma2023A&A...676A..15M} and shows gradients approaching Clump\,3 from the left and the right side of $\sim$\,-0.42 and -0.09\,km\,s$\rm ^{-1}$\,pc$\rm ^{-1}$, respectively. 
These values are consistent with that detected in several other HFSs, for instance, $\sim$\,0.8\,km\,s$\rm ^{-1}$\,pc$\rm ^{-1}$ in DR 21 South Filament \citep{hu2021dr}, 0.15\,-\,0.6\,km\,s$\rm ^{-1}$\,pc$\rm ^{-1}$ in SDC13 \citep{peretto2014sdc13}, and 0.43\,-\,0.45\,km\,s$\rm ^{-1}$\,pc$\rm ^{-1}$ in G323.46-0.08 \citep{Ma2023A&A...676A..15M}. 
However, F\,9 shows a much larger velocity gradient approaching Clump\,3, $\sim$\,-2.72 \,km\,s$\rm ^{-1}$\,pc$\rm ^{-1}$, which is possibly due to projection effects. 
Both F\,1 and F\,9 are transferring materials onto Clump\,3.

A typical V-shaped structure appearing around Clump\,3 is in good agreement with simulation results of \citet{kuznetsova2018kinematics} and \citet{gomez2014filaments}, and is similar to that detected in Orion \citep{hacar2017gravitational} and G323.46-0.08 \citep{Ma2023A&A...676A..15M}. This indicates that accelerated gravitational collapse is taking place there; the velocity gradients on the two sides of Clump\,3 are $\sim$\,-1.48 and 1.08\,km\,s$\rm ^{-1}$\,pc$\rm ^{-1}$ (see Fig.\ref{f1_f9_pv}). 
Such values are smaller than the theoretical values reported by \citet[4\,-\,7\,km\,s$\rm ^{-1}$\,pc$\rm ^{-1}$]{kuznetsova2018kinematics} and the observational results in Orion (5\,-\,7\,km\,s$\rm ^{-1}$\,pc$\rm ^{-1}$; \citealt{hacar2017gravitational}), but they are close to those obtained by \citet[1\,-\,3\,km\,s$\rm ^{-1}$\,pc$\rm ^{-1}$]{Ma2023A&A...676A..15M} in G323.46-0.08. These differences may result from optically thick CO lines or projection effects.

\subsubsection{The gravitational collapse of Sub-hub\,1}

\begin{figure}
    \centering  \includegraphics[width=0.5\textwidth]{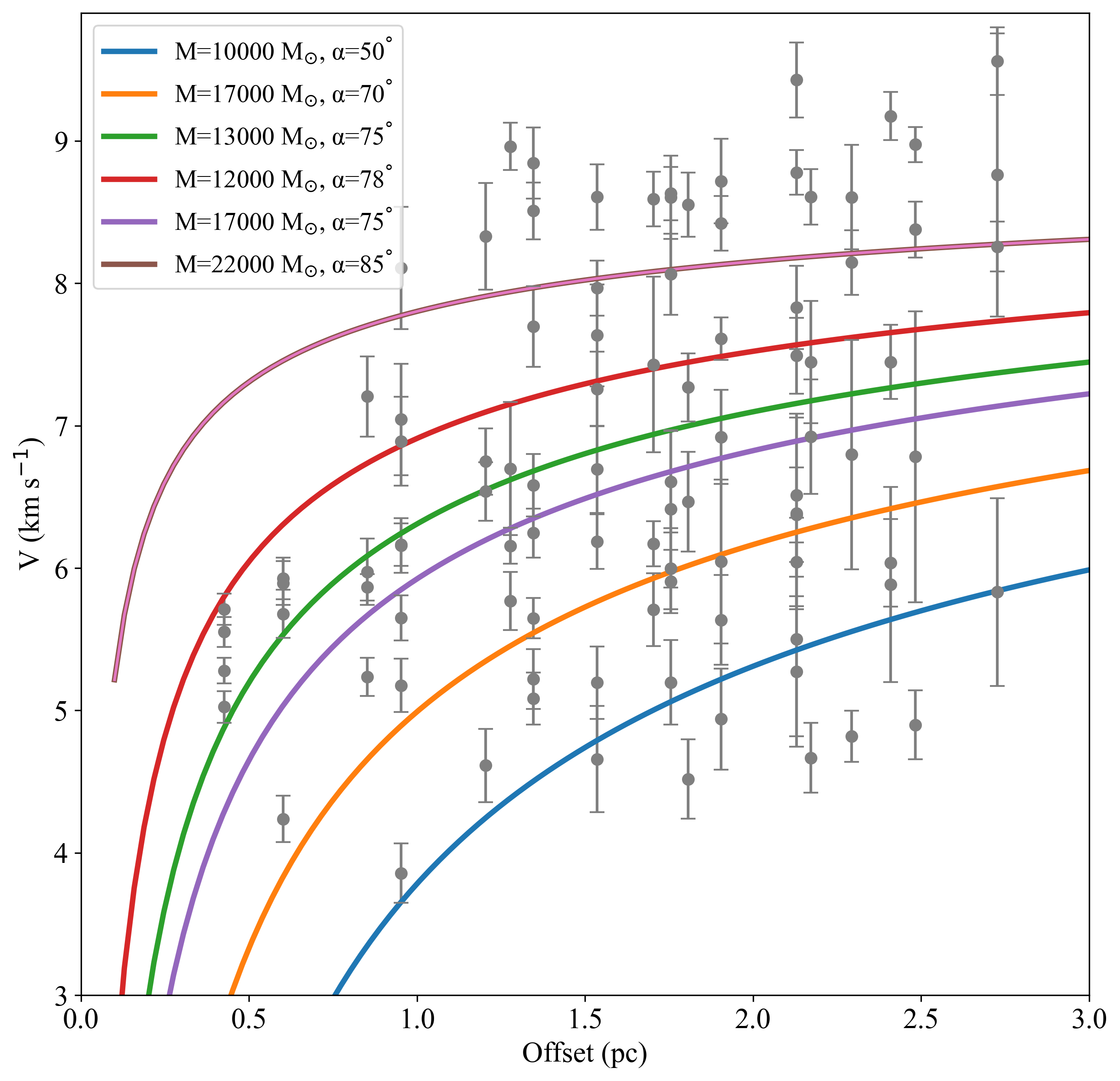}
    \caption{Gas velocity structure as a function of the distance to the center of Clump\,3 ($l_0$ = 43.23$^{\circ}$, $b_0$ = -0.05$^{\circ}$). Different fitted lines describe the expected velocity profile for a free-falling particle into potential wells with different masses, $M$, and observed at different angles, $\alpha$, following Eq.(\ref{eq.9}).}
    \label{sub_hub1_collapse}
\end{figure}

If Clump\,3 gravitationally collapses in free fall, the observed line-of-sight velocity can be described by a given impact parameter $p$ using the following relationship \citep{hacar2017gravitational}:
\begin{equation}
    V_\text{LSR}(p) = V_\text{sys,0} + V_\text{infall}(p) \cdot \cos{\alpha},
\label{eq.8}
\end{equation}
 and
\begin{equation}
    V_\text{infall}(p) = -({\frac{2GM}{R}}) ^ {\frac{1}{2}} = -({\frac{2GM}{p/\sin{\alpha}}}) ^ {\frac{1}{2}}, 
\label{eq.9}
\end{equation}
where $V_\text{sys,0}$ = 9.0 km\,s$\rm ^{-1}$ is the systemic velocity (as shown in Fig.\ref{f1_f9_pv}), and V$_\text{infall}$ is the infall velocity into a potential of mass $M$. The distance to its center, $R$, depends on the projected distance and the direction angle of the infall motion relative to the line of sight, $\alpha$.

The expected velocity profiles of free-falling particles in different potential wells and with different orientation angles are displayed in Fig.\ref{sub_hub1_collapse}. 
These model results show that a mass of 13000\,M$_{\odot}$ and a direction angle of 75$^{\circ}$ provide a better match for the observations. This model mass for Clump\,3 is nearly the same as that derived from dust emission (see Table\,\ref{tab:dense_structures}). 
The observed gradient appears to be very small for points at distances greater than $\sim$\,2\,pc, but it rapidly increases as the points approach the center of Clump\,3. 
This suggests that the motion is indeed dominated by gravity, which is consistent with theoretical results that filament collapse is slower than spheroidal collapse \citep{pon2012aspect,clarke2017filamentary}.

\subsubsection{Accretion from the filament}
We used Eq.(\ref{macc}) to estimate the mass accretion rate of the filaments in Sub-hub\,1.
The resulting mass accretion rates toward Clump\,3 along F\,1 and F\,9 are 915\,M$_{\odot}$\,Myr$\rm ^{-1}$, 410\,M$_{\odot}$\,Myr$\rm ^{-1}$, and 3260\,M$_{\odot}$\,Myr$\rm ^{-1}$, respectively. 
Therefore, a total $\sim$\,4585\,M$_{\odot}$ will be transferred onto Clump\,3 in 1\,Myr.
Such a value is much higher than the $\sim$30\,M$_{\odot}$ from \citet{kirk2013filamentary}, the  $\sim$440\,M$_{\odot}$ from \citet{yuan2017high}, the $\sim$1216\,M$_{\odot}$ from \citet{Ma2023A&A...676A..15M}, and the $\sim$3000\,M$_{\odot}$ from \citet{Sen2024arXiv240407640S}. 
Considering that Clump\,3 has a very large mass, such a high accretion rate seems to be reasonable. 
However, it should be noted that contamination from the background and foreground may result in an overestimation of the filament mass and consequently the mass accretion rate.

\subsection{The relation between the blue-shifted and red-shifted hub-filament systems}
\label{collision model}

\begin{figure}
    \centering
    \includegraphics[width=0.5\textwidth]{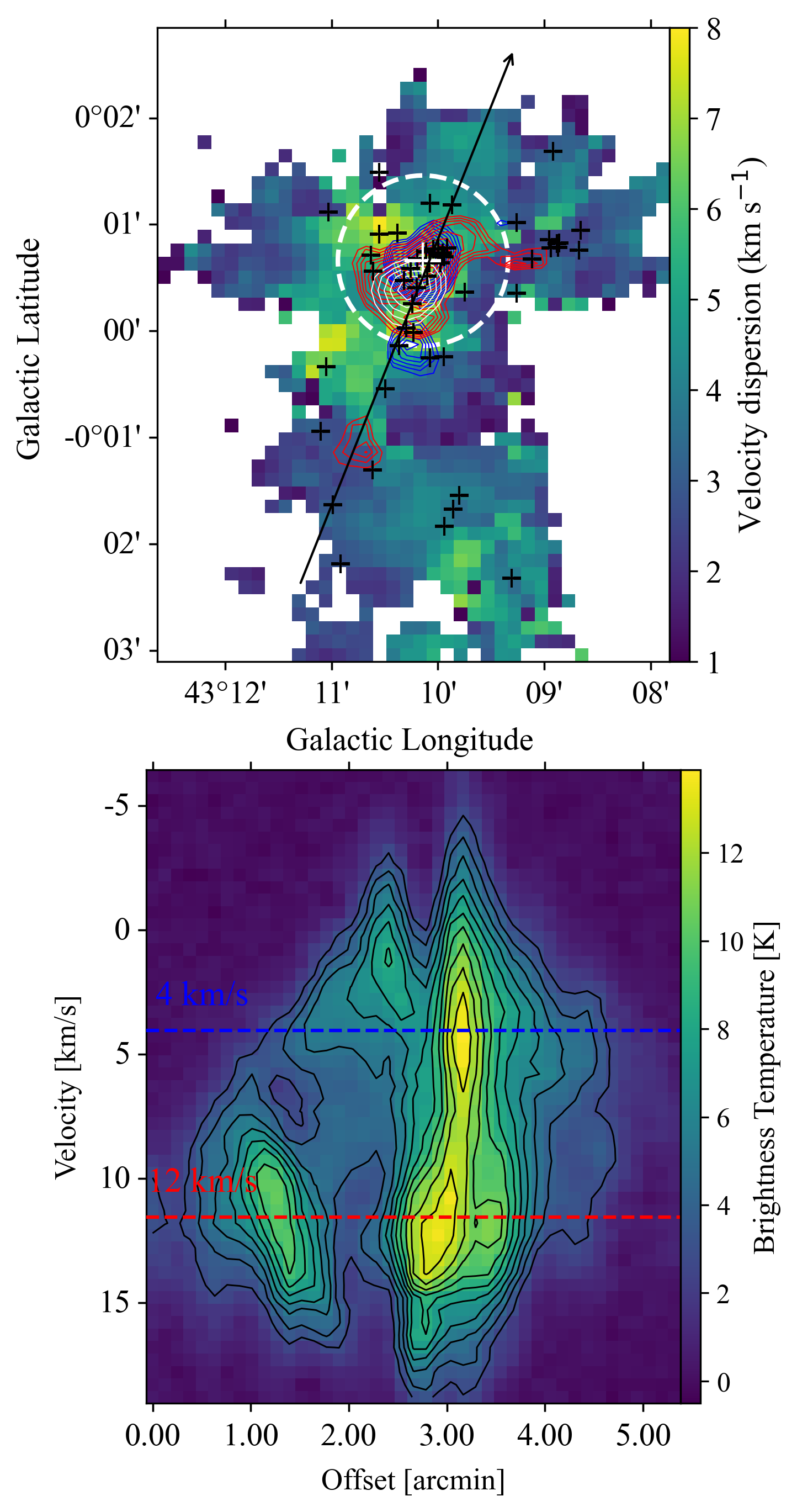}
    \caption{Velocity dispersion and P-V diagram of W49A-N. \textit{Top panel}: Velocity dispersion of $^{13}$CO\,(3-2), shown with the color map, and the velocity contours of -1\,-\,4\,km\,s$^{-1}$ and 9\,-\,14\,km\,s$^{-1}$, shown as blue and red contours, respectively. The white contours represent the strongest positions of the 1420 MHz continuum radiation (the outermost contour corresponds to 60\% of the peak intensity). The plus sign indicates the massive YSO reported by \citet{2021ApJ...923..198D}. The white circle is the position of the collision pointed out by \citet{miyawaki2022star}.
    \textit{Bottom panel}: P-V diagram along the direction of the black arrow in the top panel.}
    \label{m2 contour}
\end{figure}

The P-V diagram in Fig.\ref{m2 contour} (bottom) shows that the dense gas of the R-S HFS has a system velocity of $\sim$\,12\,km\,s$^{-1}$, and the dense gas of the B-S HFS has a velocity of $\sim$\,4\,km\,s$^{-1}$. We can see that hubs of the B-S and R-S HFSs are located nearly at the same position in W49A-N ( see the top panel of Fig.\ref{m2 contour}). There is a "bridge" connecting the B-S and R-S hubs in the P-V diagram (see the bottom panel of Fig.\ref{m2 contour}), which is usually thought as one convincing piece of evidence for the existence of cloud-cloud collision \citep{fukui2021cloud}. This suggests a head-on collision between B-S and R-S hubs may have occurred along the line of sight. Additionally, the surface density of massive YSOs reaches its maximum at the position where B-S hub overlaps with R-S hub (see the top panel of Fig.\ref{m2 contour}). \citet{miyawaki2022star} also suggest that cloud-cloud collision occurs at the same position based on the high angular resolution observations of CS and SiO (see the top panel of Fig.\ref{m2 contour}). 

It should be noted that our current results cannot discard the possibility of other mechanisms, such as the feedback of nearby H\,{\scriptsize II} regions \citep{peng2010w49a} or the localized gravitational collapse \citep{galvan2013muscle}.

%-----------------------------------------------------------------
\section{Conclusions}
\label{Conclusions}
Using CO\,(3-2) emission lines, RRLs, and infrared and radio continuum data, we have studied the structure and kinematics of the GMC W49A on a large scale. Our main conclusions are as follows.

\begin{enumerate}
\item W49A consists of two HFSs on the line of sight; they have different systemic velocities and are spatially separated by an intervening H\,{\scriptsize II} region.
The R-S HFS includes a hub at a systemic velocity of $\sim$\,12\,km\,s$\rm ^{-1}$ and seven associated filaments (F\,2, F\,4, F\,5, F\,6, F\,7, F\,8, and F\,10). The B-S HFS includes a hub at a velocity of $\sim$\,4\,km\,s$\rm ^{-1}$ and four associated filaments (F\,1, F\,2, F\,3, and F\,9). 

\item There are clear velocity gradients in the B-S HFS filaments F\,1 ($\sim$\,-0.21\,km\,s$\rm ^{-1}$\,pc$\rm ^{-1}$ ), F\,2 ($\sim$\,0.30\,km\,s$\rm ^{-1}$\,pc$\rm ^{-1}$ ), and F\,3 ($\sim$\,-0.17\,km\,s$\rm ^{-1}$\,pc$\rm ^{-1}$ ) and the R-S HFS filaments F\,4 ($\sim$\,0.22\,km\,s$\rm ^{-1}$\,pc$\rm ^{-1}$), F\,5 ($\sim$\,-0.27\,km\,s$\rm ^{-1}$\,pc$\rm ^{-1}$), F\,6 ($\sim$\,0.39\,km\,s$\rm ^{-1}$\,pc$\rm ^{-1}$), F\,7 ($\sim$\,0.31\,km\,s$\rm ^{-1}$\,pc$\rm ^{-1}$), F\,8 ($\sim$\,0.27\,km\,s$\rm ^{-1}$\,pc$\rm ^{-1}$), and F\,10 ($\sim$\,0.38\,km\,s$\rm ^{-1}$\,pc$\rm ^{-1}$). 
The B-S hub shows a centripetal velocity gradient from $\sim$\,5.0 to 1.5\,km\,s$\rm ^{-1}$ from the edge of the hub to its center; it is probably gravitationally collapsing. The R-S hub also shows a centripetal velocity gradient from $\sim$\,10 to 12\,km\,s$\rm ^{-1}$ from the edge of the hub to its center. 
    
\item The absolute values of the mass accretion rates along the filaments vary from $\sim$\,8.74$(\pm1.14)\times$10$^{-4}$ to $\sim$\,8.17$(\pm1.06)\times$10$^{-3}$\,\text{M}$_\odot$\,year$^{-1}$. This indicates that there is still significant material transport into both the B-S and R-S HFSs of W49A.
\item We conducted a detailed analysis of Sub-hub\,1 (Clump\,3) in W49A and confirm it is accreting materials through filaments and is undergoing gravitational collapse.    
In addition, filaments associated with both the B-S and R-S HFSs also show evidence of local star formation, as evidenced by the V-shaped structures around clumps in their P-V diagrams. 
\item  The separation of the B-S and R-S HFSs and the H\,{\scriptsize II} region in W49A in velocity space suggests a familiar scenario of sequential star formation along the central filament structures.
After the central hub evolves into an H\,{\scriptsize II} region, the two next clumps on either side of the main filament structure experience gravitational collapse and initiate star formation.
\end{enumerate}

%----------------------------------------------------------------- 
\begin{acknowledgements}

This work was mainly supported by the National Key R$\&$D Program of China under grant No.2022YFA1603103 and the National Natural Science foundation of China (NSFC) under grant No.12373029. It was also partially supported by the National Key R$\&$D Program of China under grant No.2023YFA1608002, the NSFC under grant Nos. 12173075, and 12103082, the Natural Science Foundation of Xinjiang Uygur Autonomous Region under grant Nos. 2022D01E06, 2022D01A359, 2022D01A362, and 2023D01A11, the Tianshan Talent Program of Xinjiang Uygur Autonomous Region under grant No. 2022TSYCLJ0005, Tianchi Talents Program of Xinjiang Uygur Autonomous Region, the Chinese Academy of Sciences (CAS) “Light of West China” Program under Grant Nos. 2020-XBQNXZ-017, 2021-XBQNXZ-028, 2022-XBJCTD-003 and xbzg-zdsys-202212, the Regional Collaborative Innovation Project of XinJiang Uyghur Autonomous Region under grant No.2022E01050, and the Science Committee of the Ministry of Science and Higher Education of the Republic of Kazakhstan grant No. AP13067768. 
WAB has been supported by the Chinese Academy of Sciences President International Fellowship Initiative by Grant No. 2023VMA0030.

\end{acknowledgements}

\bibliographystyle{aa} % style aa.bst
\bibliography{template}

\begin{thebibliography}{74}
\expandafter\ifx\csname natexlab\endcsname\relax\def\natexlab#1{#1}\fi

\bibitem[{{Andr{\'e}} {et~al.}(2014){Andr{\'e}}, {Di Francesco}, {Ward-Thompson}, {Inutsuka}, {Pudritz}, \& {Pineda}}]{2014prpl.conf...27A}
{Andr{\'e}}, P., {Di Francesco}, J., {Ward-Thompson}, D., {et~al.} 2014, in Protostars and Planets VI, ed. H.~{Beuther}, R.~S. {Klessen}, C.~P. {Dullemond}, \& T.~{Henning}, 27--51

\bibitem[{{Andr{\'e}} {et~al.}(2010){Andr{\'e}}, {Men'shchikov}, {Bontemps}, {K{\"o}nyves}, {Motte}, {Schneider}, {Didelon}, {Minier}, {Saraceno}, {Ward-Thompson}, {di Francesco}, {White}, {Molinari}, {Testi}, {Abergel}, {Griffin}, {Henning}, {Royer}, {Mer{\'\i}n}, {Vavrek}, {Attard}, {Arzoumanian}, {Wilson}, {Ade}, {Aussel}, {Baluteau}, {Benedettini}, {Bernard}, {Blommaert}, {Cambr{\'e}sy}, {Cox}, {di Giorgio}, {Hargrave}, {Hennemann}, {Huang}, {Kirk}, {Krause}, {Launhardt}, {Leeks}, {Le Pennec}, {Li}, {Martin}, {Maury}, {Olofsson}, {Omont}, {Peretto}, {Pezzuto}, {Prusti}, {Roussel}, {Russeil}, {Sauvage}, {Sibthorpe}, {Sicilia-Aguilar}, {Spinoglio}, {Waelkens}, {Woodcraft}, \& {Zavagno}}]{2010A&A...518L.102A}
{Andr{\'e}}, P., {Men'shchikov}, A., {Bontemps}, S., {et~al.} 2010, \aap, 518, L102

\bibitem[{{Aniano} {et~al.}(2011){Aniano}, {Draine}, {Gordon}, \& {Sandstrom}}]{aniano2011common}
{Aniano}, G., {Draine}, B.~T., {Gordon}, K.~D., \& {Sandstrom}, K. 2011, \pasp, 123, 1218

\bibitem[{{Benjamin} {et~al.}(2003){Benjamin}, {Churchwell}, {Babler}, {Bania}, {Clemens}, {Cohen}, {Dickey}, {Indebetouw}, {Jackson}, {Kobulnicky}, {Lazarian}, {Marston}, {Mathis}, {Meade}, {Seager}, {Stolovy}, {Watson}, {Whitney}, {Wolff}, \& {Wolfire}}]{benjamin2003glimpse}
{Benjamin}, R.~A., {Churchwell}, E., {Babler}, B.~L., {et~al.} 2003, \pasp, 115, 953

\bibitem[{{Beuther} {et~al.}(2016){Beuther}, {Bihr}, {Rugel}, {Johnston}, {Wang}, {Walter}, {Brunthaler}, {Walsh}, {Ott}, {Stil}, {Henning}, {Schierhuber}, {Kainulainen}, {Heyer}, {Goldsmith}, {Anderson}, {Longmore}, {Klessen}, {Glover}, {Urquhart}, {Plume}, {Ragan}, {Schneider}, {McClure-Griffiths}, {Menten}, {Smith}, {Roy}, {Shanahan}, {Nguyen-Luong}, \& {Bigiel}}]{beuther2016hi}
{Beuther}, H., {Bihr}, S., {Rugel}, M., {et~al.} 2016, \aap, 595, A32

\bibitem[{{Bonnell} {et~al.}(2003){Bonnell}, {Bate}, \& {Vine}}]{bonnell2003hierarchical}
{Bonnell}, I.~A., {Bate}, M.~R., \& {Vine}, S.~G. 2003, \mnras, 343, 413

\bibitem[{{Clarke} {et~al.}(2017){Clarke}, {Whitworth}, {Duarte-Cabral}, \& {Hubber}}]{clarke2017filamentary}
{Clarke}, S.~D., {Whitworth}, A.~P., {Duarte-Cabral}, A., \& {Hubber}, D.~A. 2017, \mnras, 468, 2489

\bibitem[{{De Buizer} {et~al.}(2021{\natexlab{a}}){De Buizer}, {Lim}, {Liu}, {Karnath}, \& {Radomski}}]{de2021surveying}
{De Buizer}, J.~M., {Lim}, W., {Liu}, M., {Karnath}, N., \& {Radomski}, J.~T. 2021{\natexlab{a}}, \apj, 923, 198

\bibitem[{{De Buizer} {et~al.}(2021{\natexlab{b}}){De Buizer}, {Lim}, {Liu}, {Karnath}, \& {Radomski}}]{2021ApJ...923..198D}
{De Buizer}, J.~M., {Lim}, W., {Liu}, M., {Karnath}, N., \& {Radomski}, J.~T. 2021{\natexlab{b}}, \apj, 923, 198

\bibitem[{{De Pree} {et~al.}(1997){De Pree}, {Mehringer}, \& {Goss}}]{de1997multifrequency}
{De Pree}, C.~G., {Mehringer}, D.~M., \& {Goss}, W.~M. 1997, \apj, 482, 307

\bibitem[{{Dempsey} {et~al.}(2013){Dempsey}, {Thomas}, \& {Currie}}]{dempsey2013co}
{Dempsey}, J.~T., {Thomas}, H.~S., \& {Currie}, M.~J. 2013, \apjs, 209, 8

\bibitem[{{Dewangan} {et~al.}(2020){Dewangan}, {Ojha}, {Sharma}, {Palacio}, {Bhadari}, \& {Das}}]{dewangan2020new}
{Dewangan}, L.~K., {Ojha}, D.~K., {Sharma}, S., {et~al.} 2020, \apj, 903, 13

\bibitem[{{Fukui} {et~al.}(2021){Fukui}, {Habe}, {Inoue}, {Enokiya}, \& {Tachihara}}]{fukui2021cloud}
{Fukui}, Y., {Habe}, A., {Inoue}, T., {Enokiya}, R., \& {Tachihara}, K. 2021, \pasj, 73, S1

\bibitem[{{Galv{\'a}n-Madrid} {et~al.}(2013){Galv{\'a}n-Madrid}, {Liu}, {Zhang}, {Pineda}, {Peng}, {Zhang}, {Keto}, {Ho}, {Rodr{\'\i}guez}, {Zapata}, {Peters}, \& {De Pree}}]{galvan2013muscle}
{Galv{\'a}n-Madrid}, R., {Liu}, H.~B., {Zhang}, Z.~Y., {et~al.} 2013, \apj, 779, 121

\bibitem[{{G{\'o}mez} \& {V{\'a}zquez-Semadeni}(2014)}]{gomez2014filaments}
{G{\'o}mez}, G.~C. \& {V{\'a}zquez-Semadeni}, E. 2014, \apj, 791, 124

\bibitem[{{Griffin} {et~al.}(2010){Griffin}, {Abergel}, {Abreu}, {Ade}, {Andr{\'e}}, {Augueres}, {Babbedge}, {Bae}, {Baillie}, {Baluteau}, {Barlow}, {Bendo}, {Benielli}, {Bock}, {Bonhomme}, {Brisbin}, {Brockley-Blatt}, {Caldwell}, {Cara}, {Castro-Rodriguez}, {Cerulli}, {Chanial}, {Chen}, {Clark}, {Clements}, {Clerc}, {Coker}, {Communal}, {Conversi}, {Cox}, {Crumb}, {Cunningham}, {Daly}, {Davis}, {de Antoni}, {Delderfield}, {Devin}, {di Giorgio}, {Didschuns}, {Dohlen}, {Donati}, {Dowell}, {Dowell}, {Duband}, {Dumaye}, {Emery}, {Ferlet}, {Ferrand}, {Fontignie}, {Fox}, {Franceschini}, {Frerking}, {Fulton}, {Garcia}, {Gastaud}, {Gear}, {Glenn}, {Goizel}, {Griffin}, {Grundy}, {Guest}, {Guillemet}, {Hargrave}, {Harwit}, {Hastings}, {Hatziminaoglou}, {Herman}, {Hinde}, {Hristov}, {Huang}, {Imhof}, {Isaak}, {Israelsson}, {Ivison}, {Jennings}, {Kiernan}, {King}, {Lange}, {Latter}, {Laurent}, {Laurent}, {Leeks}, {Lellouch}, {Levenson}, {Li}, {Li}, {Lilienthal}, {Lim}, {Liu}, {Lu}, {Madden}, {Mainetti}, {Marliani},
  {McKay}, {Mercier}, {Molinari}, {Morris}, {Moseley}, {Mulder}, {Mur}, {Naylor}, {Nguyen}, {O'Halloran}, {Oliver}, {Olofsson}, {Olofsson}, {Orfei}, {Page}, {Pain}, {Panuzzo}, {Papageorgiou}, {Parks}, {Parr-Burman}, {Pearce}, {Pearson}, {P{\'e}rez-Fournon}, {Pinsard}, {Pisano}, {Podosek}, {Pohlen}, {Polehampton}, {Pouliquen}, {Rigopoulou}, {Rizzo}, {Roseboom}, {Roussel}, {Rowan-Robinson}, {Rownd}, {Saraceno}, {Sauvage}, {Savage}, {Savini}, {Sawyer}, {Scharmberg}, {Schmitt}, {Schneider}, {Schulz}, {Schwartz}, {Shafer}, {Shupe}, {Sibthorpe}, {Sidher}, {Smith}, {Smith}, {Smith}, {Spencer}, {Stobie}, {Sudiwala}, {Sukhatme}, {Surace}, {Stevens}, {Swinyard}, {Trichas}, {Tourette}, {Triou}, {Tseng}, {Tucker}, {Turner}, {Vaccari}, {Valtchanov}, {Vigroux}, {Virique}, {Voellmer}, {Walker}, {Ward}, {Waskett}, {Weilert}, {Wesson}, {White}, {Whitehouse}, {Wilson}, {Winter}, {Woodcraft}, {Wright}, {Xu}, {Zavagno}, {Zemcov}, {Zhang}, \& {Zonca}}]{griffin2010herschel}
{Griffin}, M.~J., {Abergel}, A., {Abreu}, A., {et~al.} 2010, \aap, 518, L3

\bibitem[{{Hacar} {et~al.}(2017){Hacar}, {Alves}, {Tafalla}, \& {Goicoechea}}]{hacar2017gravitational}
{Hacar}, A., {Alves}, J., {Tafalla}, M., \& {Goicoechea}, J.~R. 2017, \aap, 602, L2

\bibitem[{{He} {et~al.}(2023){He}, {Liu}, {Tang}, {Qin}, {Zhou}, {Esimbek}, {Pan}, {Li}, {Zhao}, {Ji}, \& {Komesh}}]{he2023investigating}
{He}, Y.-X., {Liu}, H.-L., {Tang}, X.-D., {et~al.} 2023, \apj, 957, 61

\bibitem[{{Homeier} \& {Alves}(2005)}]{homeier2005massive}
{Homeier}, N.~L. \& {Alves}, J. 2005, \aap, 430, 481

\bibitem[{{Hu} {et~al.}(2021){Hu}, {Qiu}, {Cao}, {Liu}, {Wang}, {Li}, {Shen}, {Li}, {Wang}, {Li}, \& {Dong}}]{hu2021dr}
{Hu}, B., {Qiu}, K., {Cao}, Y., {et~al.} 2021, \apj, 908, 70

\bibitem[{{Kauffmann} {et~al.}(2008){Kauffmann}, {Bertoldi}, {Bourke}, {Evans}, \& {Lee}}]{kauffmann2008mambo}
{Kauffmann}, J., {Bertoldi}, F., {Bourke}, T.~L., {Evans}, N.~J., I., \& {Lee}, C.~W. 2008, \aap, 487, 993

\bibitem[{{Kauffmann} \& {Pillai}(2010)}]{kauffmann2010many}
{Kauffmann}, J. \& {Pillai}, T. 2010, \apjl, 723, L7

\bibitem[{{Kirk} {et~al.}(2013){Kirk}, {Myers}, {Bourke}, {Gutermuth}, {Hedden}, \& {Wilson}}]{kirk2013filamentary}
{Kirk}, H., {Myers}, P.~C., {Bourke}, T.~L., {et~al.} 2013, \apj, 766, 115

\bibitem[{{Koch} \& {Rosolowsky}(2015)}]{2015MNRAS.452.3435K}
{Koch}, E.~W. \& {Rosolowsky}, E.~W. 2015, \mnras, 452, 3435

\bibitem[{{Kuhn} {et~al.}(2021){Kuhn}, {de Souza}, {Krone-Martins}, {Castro-Ginard}, {Ishida}, {Povich}, {Hillenbrand}, \& {COIN Collaboration}}]{kuhn2021spicy}
{Kuhn}, M.~A., {de Souza}, R.~S., {Krone-Martins}, A., {et~al.} 2021, \apjs, 254, 33

\bibitem[{{Kumar} {et~al.}(2020){Kumar}, {Palmeirim}, {Arzoumanian}, \& {Inutsuka}}]{kumar2020unifying}
{Kumar}, M.~S.~N., {Palmeirim}, P., {Arzoumanian}, D., \& {Inutsuka}, S.~I. 2020, \aap, 642, A87

\bibitem[{{Kuznetsova} {et~al.}(2018){Kuznetsova}, {Hartmann}, \& {Ballesteros-Paredes}}]{kuznetsova2018kinematics}
{Kuznetsova}, A., {Hartmann}, L., \& {Ballesteros-Paredes}, J. 2018, \mnras, 473, 2372

\bibitem[{{Li} {et~al.}(2016){Li}, {Urquhart}, {Leurini}, {Csengeri}, {Wyrowski}, {Menten}, \& {Schuller}}]{li2016atlasgal}
{Li}, G.-X., {Urquhart}, J.~S., {Leurini}, S., {et~al.} 2016, \aap, 591, A5

\bibitem[{{Lin} {et~al.}(2016){Lin}, {Liu}, {Li}, {Zhang}, {Ginsburg}, {Pineda}, {Qian}, {Galv{\'a}n-Madrid}, {McLeod}, {Rosolowsky}, {Dale}, {Immer}, {Koch}, {Longmore}, {Walker}, \& {Testi}}]{lin2016cloud}
{Lin}, Y., {Liu}, H.~B., {Li}, D., {et~al.} 2016, \apj, 828, 32

\bibitem[{{Liu} {et~al.}(2023){Liu}, {Tej}, {Liu}, {Sanhueza}, {Qin}, {He}, {Goldsmith}, {Garay}, {Pan}, {Morii}, {Li}, {Stutz}, {Tatematsu}, {Xu}, {Bronfman}, {Saha}, {Issac}, {Baug}, {Toth}, {Dewangan}, {Wang}, {Zhou}, {Lee}, {Yang}, {Luo}, {Shen}, {Zhang}, {Wu}, {Ren}, {Liu}, {Soam}, {Zhang}, \& {Luo}}]{liu2023evidence}
{Liu}, H.-L., {Tej}, A., {Liu}, T., {et~al.} 2023, \mnras, 522, 3719

\bibitem[{{Ma} {et~al.}(2023){Ma}, {Zhou}, {Esimbek}, {Baan}, {Li}, {Tang}, {He}, {Ji}, {Zhou}, {Wu}, {Tursun}, \& {Komesh}}]{Ma2023A&A...676A..15M}
{Ma}, Y., {Zhou}, J., {Esimbek}, J., {et~al.} 2023, \aap, 676, A15

\bibitem[{{Mattern} {et~al.}(2018){Mattern}, {Kainulainen}, {Zhang}, \& {Beuther}}]{2018A&A...616A..78M}
{Mattern}, M., {Kainulainen}, J., {Zhang}, M., \& {Beuther}, H. 2018, \aap, 616, A78

\bibitem[{{Mazumdar} {et~al.}(2021){Mazumdar}, {Wyrowski}, {Urquhart}, {Colombo}, {Menten}, {Neupane}, \& {Thompson}}]{mazumdar2021high}
{Mazumdar}, P., {Wyrowski}, F., {Urquhart}, J.~S., {et~al.} 2021, \aap, 656, A101

\bibitem[{{Miyawaki} {et~al.}(1986){Miyawaki}, {Hayashi}, \& {Hasegawa}}]{miyawaki1986structure}
{Miyawaki}, R., {Hayashi}, M., \& {Hasegawa}, T. 1986, \apj, 305, 353

\bibitem[{{Miyawaki} {et~al.}(2009){Miyawaki}, {Hayashi}, \& {Hasegawa}}]{miyawaki2009large}
{Miyawaki}, R., {Hayashi}, M., \& {Hasegawa}, T. 2009, \pasj, 61, 39

\bibitem[{{Miyawaki} {et~al.}(2022){Miyawaki}, {Hayashi}, \& {Hasegawa}}]{miyawaki2022star}
{Miyawaki}, R., {Hayashi}, M., \& {Hasegawa}, T. 2022, \pasj, 74, 128

\bibitem[{{Molinari} {et~al.}(2010{\natexlab{a}}){Molinari}, {Swinyard}, {Bally}, {Barlow}, {Bernard}, {Martin}, {Moore}, {Noriega-Crespo}, {Plume}, {Testi}, {Zavagno}, {Abergel}, {Ali}, {Anderson}, {Andr{\'e}}, {Baluteau}, {Battersby}, {Beltr{\'a}n}, {Benedettini}, {Billot}, {Blommaert}, {Bontemps}, {Boulanger}, {Brand}, {Brunt}, {Burton}, {Calzoletti}, {Carey}, {Caselli}, {Cesaroni}, {Cernicharo}, {Chakrabarti}, {Chrysostomou}, {Cohen}, {Compiegne}, {de Bernardis}, {de Gasperis}, {di Giorgio}, {Elia}, {Faustini}, {Flagey}, {Fukui}, {Fuller}, {Ganga}, {Garcia-Lario}, {Glenn}, {Goldsmith}, {Griffin}, {Hoare}, {Huang}, {Ikhenaode}, {Joblin}, {Joncas}, {Juvela}, {Kirk}, {Lagache}, {Li}, {Lim}, {Lord}, {Marengo}, {Marshall}, {Masi}, {Massi}, {Matsuura}, {Minier}, {Miville-Desch{\^e}nes}, {Montier}, {Morgan}, {Motte}, {Mottram}, {M{\"u}ller}, {Natoli}, {Neves}, {Olmi}, {Paladini}, {Paradis}, {Parsons}, {Peretto}, {Pestalozzi}, {Pezzuto}, {Piacentini}, {Piazzo}, {Polychroni}, {Pomar{\`e}s}, {Popescu}, {Reach},
  {Ristorcelli}, {Robitaille}, {Robitaille}, {Rod{\'o}n}, {Roy}, {Royer}, {Russeil}, {Saraceno}, {Sauvage}, {Schilke}, {Schisano}, {Schneider}, {Schuller}, {Schulz}, {Sibthorpe}, {Smith}, {Smith}, {Spinoglio}, {Stamatellos}, {Strafella}, {Stringfellow}, {Sturm}, {Taylor}, {Thompson}, {Traficante}, {Tuffs}, {Umana}, {Valenziano}, {Vavrek}, {Veneziani}, {Viti}, {Waelkens}, {Ward-Thompson}, {White}, {Wilcock}, {Wyrowski}, {Yorke}, \& {Zhang}}]{molinari2010clouds}
{Molinari}, S., {Swinyard}, B., {Bally}, J., {et~al.} 2010{\natexlab{a}}, \aap, 518, L100

\bibitem[{{Molinari} {et~al.}(2010{\natexlab{b}}){Molinari}, {Swinyard}, {Bally}, {Barlow}, {Bernard}, {Martin}, {Moore}, {Noriega-Crespo}, {Plume}, {Testi}, {Zavagno}, {Abergel}, {Ali}, {Andr{\'e}}, {Baluteau}, {Benedettini}, {Bern{\'e}}, {Billot}, {Blommaert}, {Bontemps}, {Boulanger}, {Brand}, {Brunt}, {Burton}, {Campeggio}, {Carey}, {Caselli}, {Cesaroni}, {Cernicharo}, {Chakrabarti}, {Chrysostomou}, {Codella}, {Cohen}, {Compiegne}, {Davis}, {de Bernardis}, {de Gasperis}, {Di Francesco}, {di Giorgio}, {Elia}, {Faustini}, {Fischera}, {Fukui}, {Fuller}, {Ganga}, {Garcia-Lario}, {Giard}, {Giardino}, {Glenn}, {Goldsmith}, {Griffin}, {Hoare}, {Huang}, {Jiang}, {Joblin}, {Joncas}, {Juvela}, {Kirk}, {Lagache}, {Li}, {Lim}, {Lord}, {Lucas}, {Maiolo}, {Marengo}, {Marshall}, {Masi}, {Massi}, {Matsuura}, {Meny}, {Minier}, {Miville-Desch{\^e}nes}, {Montier}, {Motte}, {M{\"u}ller}, {Natoli}, {Neves}, {Olmi}, {Paladini}, {Paradis}, {Pestalozzi}, {Pezzuto}, {Piacentini}, {Pomar{\`e}s}, {Popescu}, {Reach}, {Richer},
  {Ristorcelli}, {Roy}, {Royer}, {Russeil}, {Saraceno}, {Sauvage}, {Schilke}, {Schneider-Bontemps}, {Schuller}, {Schultz}, {Shepherd}, {Sibthorpe}, {Smith}, {Smith}, {Spinoglio}, {Stamatellos}, {Strafella}, {Stringfellow}, {Sturm}, {Taylor}, {Thompson}, {Tuffs}, {Umana}, {Valenziano}, {Vavrek}, {Viti}, {Waelkens}, {Ward-Thompson}, {White}, {Wyrowski}, {Yorke}, \& {Zhang}}]{molinari2010hi}
{Molinari}, S., {Swinyard}, B., {Bally}, J., {et~al.} 2010{\natexlab{b}}, \pasp, 122, 314

\bibitem[{{Mookerjea} {et~al.}(2023){Mookerjea}, {Veena}, {G{\"u}sten}, {Wyrowski}, \& {Lasrado}}]{mookerjea2023spiral}
{Mookerjea}, B., {Veena}, V.~S., {G{\"u}sten}, R., {Wyrowski}, F., \& {Lasrado}, A. 2023, \mnras, 520, 2517

\bibitem[{{Mufson} \& {Liszt}(1977)}]{mufson1977structure}
{Mufson}, S.~L. \& {Liszt}, H.~S. 1977, \apj, 212, 664

\bibitem[{{Myers}(2009)}]{myers2009filamentary}
{Myers}, P.~C. 2009, \apj, 700, 1609

\bibitem[{{Ossenkopf} \& {Henning}(1994)}]{ossenkopf1994dust}
{Ossenkopf}, V. \& {Henning}, T. 1994, \aap, 291, 943

\bibitem[{{Palmeirim} {et~al.}(2013){Palmeirim}, {Andr{\'e}}, {Kirk}, {Ward-Thompson}, {Arzoumanian}, {K{\"o}nyves}, {Didelon}, {Schneider}, {Benedettini}, {Bontemps}, {Di Francesco}, {Elia}, {Griffin}, {Hennemann}, {Hill}, {Martin}, {Men'shchikov}, {Molinari}, {Motte}, {Nguyen Luong}, {Nutter}, {Peretto}, {Pezzuto}, {Roy}, {Rygl}, {Spinoglio}, \& {White}}]{palmeirim2013herschel}
{Palmeirim}, P., {Andr{\'e}}, P., {Kirk}, J., {et~al.} 2013, \aap, 550, A38

\bibitem[{{Paron} {et~al.}(2018){Paron}, {Areal}, \& {Ortega}}]{paron2018mapping}
{Paron}, S., {Areal}, M.~B., \& {Ortega}, M.~E. 2018, \aap, 617, A14

\bibitem[{{Peng} {et~al.}(2010){Peng}, {Wyrowski}, {van der Tak}, {Menten}, \& {Walmsley}}]{peng2010w49a}
{Peng}, T.~C., {Wyrowski}, F., {van der Tak}, F.~F.~S., {Menten}, K.~M., \& {Walmsley}, C.~M. 2010, \aap, 520, A84

\bibitem[{{Peretto} {et~al.}(2014){Peretto}, {Fuller}, {Andr{\'e}}, {Arzoumanian}, {Rivilla}, {Bardeau}, {Duarte Puertas}, {Guzman Fernandez}, {Lenfestey}, {Li}, {Olguin}, {R{\"o}ck}, {de Villiers}, \& {Williams}}]{peretto2014sdc13}
{Peretto}, N., {Fuller}, G.~A., {Andr{\'e}}, P., {et~al.} 2014, \aap, 561, A83

\bibitem[{{Peretto} {et~al.}(2013){Peretto}, {Fuller}, {Duarte-Cabral}, {Avison}, {Hennebelle}, {Pineda}, {Andr{\'e}}, {Bontemps}, {Motte}, {Schneider}, \& {Molinari}}]{peretto2013global}
{Peretto}, N., {Fuller}, G.~A., {Duarte-Cabral}, A., {et~al.} 2013, \aap, 555, A112

\bibitem[{{Pilbratt} {et~al.}(2010){Pilbratt}, {Riedinger}, {Passvogel}, {Crone}, {Doyle}, {Gageur}, {Heras}, {Jewell}, {Metcalfe}, {Ott}, \& {Schmidt}}]{pilbratt2010herschel}
{Pilbratt}, G.~L., {Riedinger}, J.~R., {Passvogel}, T., {et~al.} 2010, \aap, 518, L1

\bibitem[{{Poglitsch} {et~al.}(2010){Poglitsch}, {Waelkens}, {Bauer}, {Cepa}, {Feuchtgruber}, {Henning}, {van Hoof}, {Kerschbaum}, {Krause}, {Renotte}, {Saraceno}, \& {Vandenbussche}}]{poglitsch2010herschel}
{Poglitsch}, A., {Waelkens}, C., {Bauer}, O.~H., {et~al.} 2010, in 38th COSPAR Scientific Assembly, Vol.~38, 13

\bibitem[{{Pon} {et~al.}(2012){Pon}, {Toal{\'a}}, {Johnstone}, {V{\'a}zquez-Semadeni}, {Heitsch}, \& {G{\'o}mez}}]{pon2012aspect}
{Pon}, A., {Toal{\'a}}, J.~A., {Johnstone}, D., {et~al.} 2012, \apj, 756, 145

\bibitem[{{Rieke} {et~al.}(2004){Rieke}, {Young}, {Engelbracht}, {Kelly}, {Low}, {Haller}, {Beeman}, {Gordon}, {Stansberry}, {Misselt}, {Cadien}, {Morrison}, {Rivlis}, {Latter}, {Noriega-Crespo}, {Padgett}, {Stapelfeldt}, {Hines}, {Egami}, {Muzerolle}, {Alonso-Herrero}, {Blaylock}, {Dole}, {Hinz}, {Le Floc'h}, {Papovich}, {P{\'e}rez-Gonz{\'a}lez}, {Smith}, {Su}, {Bennett}, {Frayer}, {Henderson}, {Lu}, {Masci}, {Pesenson}, {Rebull}, {Rho}, {Keene}, {Stolovy}, {Wachter}, {Wheaton}, {Werner}, \& {Richards}}]{rieke2004multiband}
{Rieke}, G.~H., {Young}, E.~T., {Engelbracht}, C.~W., {et~al.} 2004, \apjs, 154, 25

\bibitem[{{Rigby} {et~al.}(2019){Rigby}, {Moore}, {Eden}, {Urquhart}, {Ragan}, {Peretto}, {Plume}, {Thompson}, {Currie}, \& {Park}}]{rigby2019chimps}
{Rigby}, A.~J., {Moore}, T.~J.~T., {Eden}, D.~J., {et~al.} 2019, \aap, 632, A58

\bibitem[{{Rigby} {et~al.}(2016){Rigby}, {Moore}, {Plume}, {Eden}, {Urquhart}, {Thompson}, {Mottram}, {Brunt}, {Butner}, {Dempsey}, {Gibson}, {Hatchell}, {Jenness}, {Kuno}, {Longmore}, {Morgan}, {Polychroni}, {Thomas}, {White}, \& {Zhu}}]{rigby2016chimps}
{Rigby}, A.~J., {Moore}, T.~J.~T., {Plume}, R., {et~al.} 2016, \mnras, 456, 2885

\bibitem[{{Saral} {et~al.}(2015){Saral}, {Hora}, {Willis}, {Koenig}, {Gutermuth}, \& {Saygac}}]{saral2015young}
{Saral}, G., {Hora}, J.~L., {Willis}, S.~E., {et~al.} 2015, \apj, 813, 25

\bibitem[{{Schisano} {et~al.}(2020){Schisano}, {Molinari}, {Elia}, {Benedettini}, {Olmi}, {Pezzuto}, {Traficante}, {Brescia}, {Cavuoti}, {di Giorgio}, {Liu}, {Moore}, {Noriega-Crespo}, {Riccio}, {Baldeschi}, {Becciani}, {Peretto}, {Merello}, {Vitello}, {Zavagno}, {Beltr{\'a}n}, {Cambr{\'e}sy}, {Eden}, {Li Causi}, {Molinaro}, {Palmeirim}, {Sciacca}, {Testi}, {Umana}, \& {Whitworth}}]{2020MNRAS.492.5420S}
{Schisano}, E., {Molinari}, S., {Elia}, D., {et~al.} 2020, \mnras, 492, 5420

\bibitem[{{Sen} {et~al.}(2024){Sen}, {Mookerjea}, {Guesten}, {Wyrowski}, \& {Ishwara Chandra}}]{Sen2024arXiv240407640S}
{Sen}, S., {Mookerjea}, B., {Guesten}, R., {Wyrowski}, F., \& {Ishwara Chandra}, C.~H. 2024, arXiv e-prints, arXiv:2404.07640

\bibitem[{{Serabyn} {et~al.}(1993){Serabyn}, {Guesten}, \& {Schulz}}]{serabyn1993fragmentation}
{Serabyn}, E., {Guesten}, R., \& {Schulz}, A. 1993, \apj, 413, 571

\bibitem[{{Seshadri} {et~al.}(2024){Seshadri}, {Vig}, {Ghosh}, \& {Ojha}}]{Seshadri2024MNRAS.527.4244S}
{Seshadri}, A., {Vig}, S., {Ghosh}, S.~K., \& {Ojha}, D.~K. 2024, \mnras, 527, 4244

\bibitem[{{Simon} {et~al.}(2001){Simon}, {Jackson}, {Clemens}, {Bania}, \& {Heyer}}]{simon2001structure}
{Simon}, R., {Jackson}, J.~M., {Clemens}, D.~P., {Bania}, T.~M., \& {Heyer}, M.~H. 2001, \apj, 551, 747

\bibitem[{{Smith} {et~al.}(2009){Smith}, {Longmore}, \& {Bonnell}}]{smith2009simultaneous}
{Smith}, R.~J., {Longmore}, S., \& {Bonnell}, I. 2009, \mnras, 400, 1775

\bibitem[{{Trevi{\~n}o-Morales} {et~al.}(2019){Trevi{\~n}o-Morales}, {Fuente}, {S{\'a}nchez-Monge}, {Kainulainen}, {Didelon}, {Suri}, {Schneider}, {Ballesteros-Paredes}, {Lee}, {Hennebelle}, {Pilleri}, {Gonz{\'a}lez-Garc{\'\i}a}, {Kramer}, {Garc{\'\i}a-Burillo}, {Luna}, {Goicoechea}, {Tremblin}, \& {Geen}}]{trevino2019dynamics}
{Trevi{\~n}o-Morales}, S.~P., {Fuente}, A., {S{\'a}nchez-Monge}, {\'A}., {et~al.} 2019, \aap, 629, A81

\bibitem[{{Urquhart} {et~al.}(2014){Urquhart}, {Csengeri}, {Wyrowski}, {Schuller}, {Bontemps}, {Bronfman}, {Menten}, {Walmsley}, {Contreras}, {Beuther}, {Wienen}, \& {Linz}}]{urquhart2014atlasgal}
{Urquhart}, J.~S., {Csengeri}, T., {Wyrowski}, F., {et~al.} 2014, \aap, 568, A41

\bibitem[{{Urquhart} {et~al.}(2018){Urquhart}, {K{\"o}nig}, {Giannetti}, {Leurini}, {Moore}, {Eden}, {Pillai}, {Thompson}, {Braiding}, {Burton}, {Csengeri}, {Dempsey}, {Figura}, {Froebrich}, {Menten}, {Schuller}, {Smith}, \& {Wyrowski}}]{urquhart2018atlasgal}
{Urquhart}, J.~S., {K{\"o}nig}, C., {Giannetti}, A., {et~al.} 2018, \mnras, 473, 1059

\bibitem[{{V{\'a}zquez-Semadeni} {et~al.}(2019){V{\'a}zquez-Semadeni}, {Palau}, {Ballesteros-Paredes}, {G{\'o}mez}, \& {Zamora-Avil{\'e}s}}]{vazquez2019global}
{V{\'a}zquez-Semadeni}, E., {Palau}, A., {Ballesteros-Paredes}, J., {G{\'o}mez}, G.~C., \& {Zamora-Avil{\'e}s}, M. 2019, \mnras, 490, 3061

\bibitem[{{Wang} {et~al.}(2020{\natexlab{a}}){Wang}, {Koch}, {Galv{\'a}n-Madrid}, {Lai}, {Liu}, {Lin}, \& {Pattle}}]{wang2020formation}
{Wang}, J.-W., {Koch}, P.~M., {Galv{\'a}n-Madrid}, R., {et~al.} 2020{\natexlab{a}}, \apj, 905, 158

\bibitem[{{Wang} {et~al.}(2015){Wang}, {Testi}, {Ginsburg}, {Walmsley}, {Molinari}, \& {Schisano}}]{wang2015large}
{Wang}, K., {Testi}, L., {Ginsburg}, A., {et~al.} 2015, \mnras, 450, 4043

\bibitem[{{Wang} {et~al.}(2020{\natexlab{b}}){Wang}, {Beuther}, {Rugel}, {Soler}, {Stil}, {Ott}, {Bihr}, {McClure-Griffiths}, {Anderson}, {Klessen}, {Goldsmith}, {Roy}, {Glover}, {Urquhart}, {Heyer}, {Linz}, {Smith}, {Bigiel}, {Dempsey}, \& {Henning}}]{wang2020hi}
{Wang}, Y., {Beuther}, H., {Rugel}, M.~R., {et~al.} 2020{\natexlab{b}}, \aap, 634, A83

\bibitem[{{Welch} {et~al.}(1987){Welch}, {Dreher}, {Jackson}, {Terebey}, \& {Vogel}}]{welch1987star}
{Welch}, W.~J., {Dreher}, J.~W., {Jackson}, J.~M., {Terebey}, S., \& {Vogel}, S.~N. 1987, Science, 238, 1550

\bibitem[{{Wu} {et~al.}(2016){Wu}, {Bik}, {Bestenlehner}, {Henning}, {Pasquali}, {Brandner}, \& {Stolte}}]{wu2016massive}
{Wu}, S.-W., {Bik}, A., {Bestenlehner}, J.~M., {et~al.} 2016, \aap, 589, A16

\bibitem[{{Xu} {et~al.}(2023){Xu}, {Wang}, {Liu}, {Goldsmith}, {Zhang}, {Juvela}, {Liu}, {Qin}, {Li}, {Tej}, {Garay}, {Bronfman}, {Li}, {Wu}, {G{\'o}mez}, {V{\'a}zquez-Semadeni}, {Tatematsu}, {Ren}, {Zhang}, {Toth}, {Liu}, {Yue}, {Zhang}, {Baug}, {Issac}, {Stutz}, {Liu}, {Fuller}, {Tang}, {Zhang}, {Dewangan}, {Lee}, {Zhou}, {Xie}, {Jiao}, {Wang}, {Liu}, {Luo}, {Soam}, \& {Eswaraiah}}]{xu2023atoms}
{Xu}, F.-W., {Wang}, K., {Liu}, T., {et~al.} 2023, \mnras, 520, 3259

\bibitem[{{Yang} {et~al.}(2023){Yang}, {Liu}, {Tej}, {Liu}, {Sanhueza}, {Qin}, {Lu}, {Wang}, {Pan}, {Xu}, {V{\'a}zquez-Semadeni​}, {Li}, {G{\'o}mez}, {Palau}, {Garay}, {Goldsmith}, {Juvela}, {Saha}, {Bronfman}, {Lee}, {Tatematsu}, {Dewangan}, {Zhou}, {Zhang}, {Stutz}, {Eswaraiah}, {Toth}, {Ristorcelli}, {Shen}, {Luo}, \& {Chibueze}}]{yang2023direct}
{Yang}, D., {Liu}, H.-L., {Tej}, A., {et~al.} 2023, \apj, 953, 40

\bibitem[{{Yuan} {et~al.}(2018){Yuan}, {Li}, {Wu}, {Ellingsen}, {Henkel}, {Wang}, {Liu}, {Liu}, {Zavagno}, {Ren}, \& {Huang}}]{yuan2017high}
{Yuan}, J., {Li}, J.-Z., {Wu}, Y., {et~al.} 2018, \apj, 852, 12

\bibitem[{{Zhang} {et~al.}(2013){Zhang}, {Reid}, {Menten}, {Zheng}, {Brunthaler}, {Dame}, \& {Xu}}]{zhang2013parallaxes}
{Zhang}, B., {Reid}, M.~J., {Menten}, K.~M., {et~al.} 2013, \apj, 775, 79

\bibitem[{{Zhou} {et~al.}(2022){Zhou}, {Liu}, {Evans}, {Garay}, {Goldsmith}, {G{\'o}mez}, {V{\'a}zquez-Semadeni}, {Liu}, {Stutz}, {Wang}, {Juvela}, {He}, {Li}, {Bronfman}, {Liu}, {Xu}, {Tej}, {Dewangan}, {Li}, {Zhang}, {Zhang}, {Ren}, {Tatematsu}, {Shing Li}, {Won Lee}, {Baug}, {Qin}, {Wu}, {Peng}, {Zhang}, {Liu}, {Luo}, {Ge}, {Saha}, {Chakali}, {Zhang}, {Kim}, {Ristorcelli}, {Shen}, \& {Li}}]{zhou2022atoms}
{Zhou}, J.-W., {Liu}, T., {Evans}, N.~J., {et~al.} 2022, \mnras, 514, 6038

\end{thebibliography}

\begin{appendix}

\onecolumn

\newpage
\section{Interpolating saturated pixels in \textit{Herschel} images}
\label{appendix:interpolating}
\begin{figure*}[hbt]
    \centering
    \includegraphics[width=\textwidth]{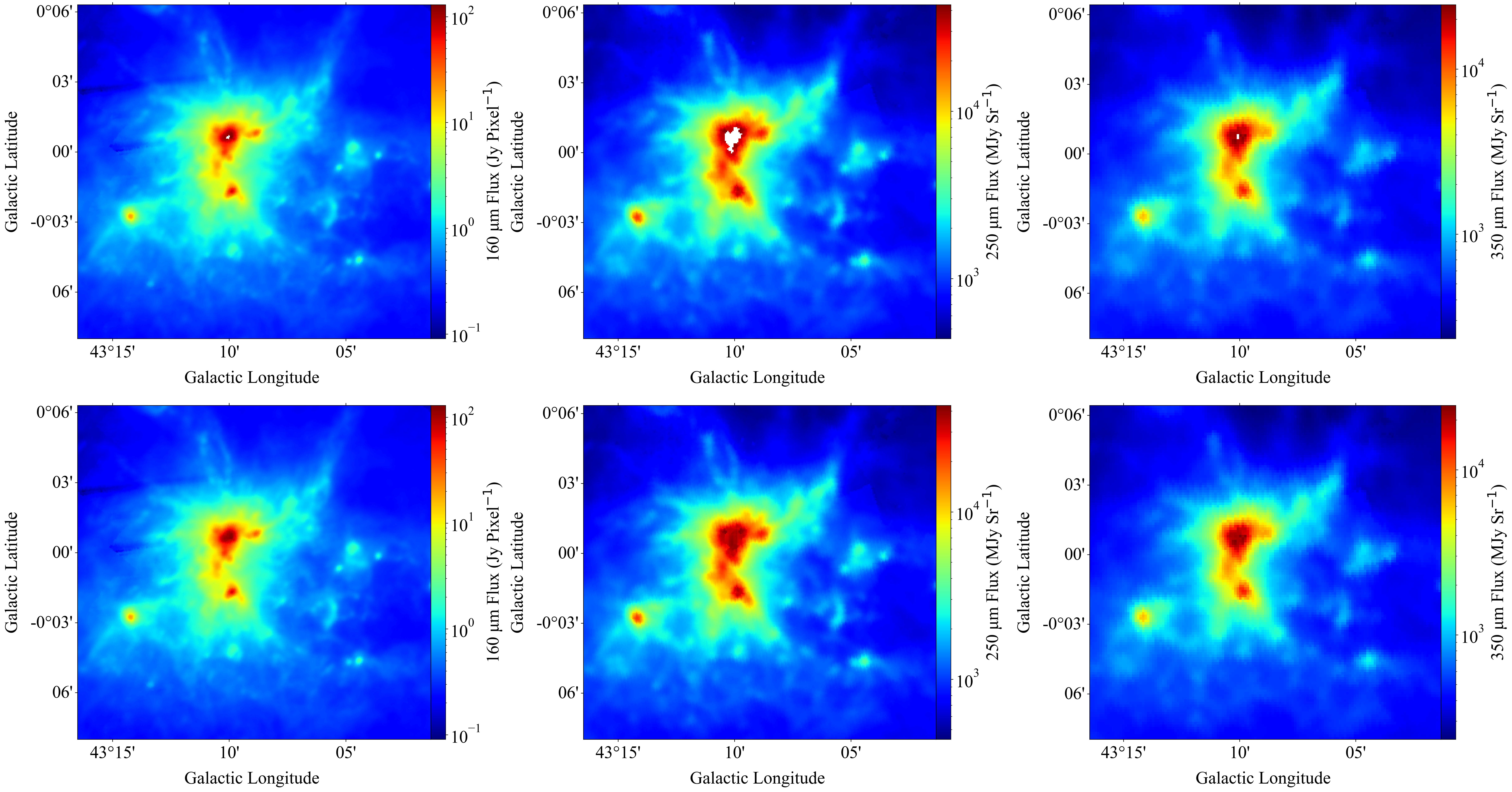}
    \caption{Original images at 160, 250, and 350\,µm in W49A (top panels) and the images after filling in the saturated regions (bottom panels).}
    \label{fig:interpolating Herschel image}
\end{figure*}

A few pixels are saturated at the center of W49-N in \textit{Herschel} images at 160, 250, and 350\,µm. Similar to \citet{lin2016cloud}, we used the interpolated values to replace those saturated pixels. Since the flux in saturated regions generally represents the peak flux of that area and decreases gradually from the center outward, to some extent, it exhibits a Gaussian distribution of radiation. Therefore, we used Gaussian fitting interpolation to recover the missed flux in the saturated regions. The interpolated images are shown in Fig.\ref{fig:interpolating Herschel image}. 

\section{Analysis of CO optical depth in W49A}
\label{appendix:optical depth}
The optical depth of CO will affect our analysis of velocity gradients in the double velocity components in W49A. This is because self-absorption in optically thick molecules can affect their measured spectra. Additionally, since it is J = 3-2, it may only trace relatively high-excitation gas. Therefore, it is necessary to discuss the optical depth of CO molecules in the W49A region here.

\begin{figure*}[hbt]
    \centering
    \includegraphics[width=\textwidth]{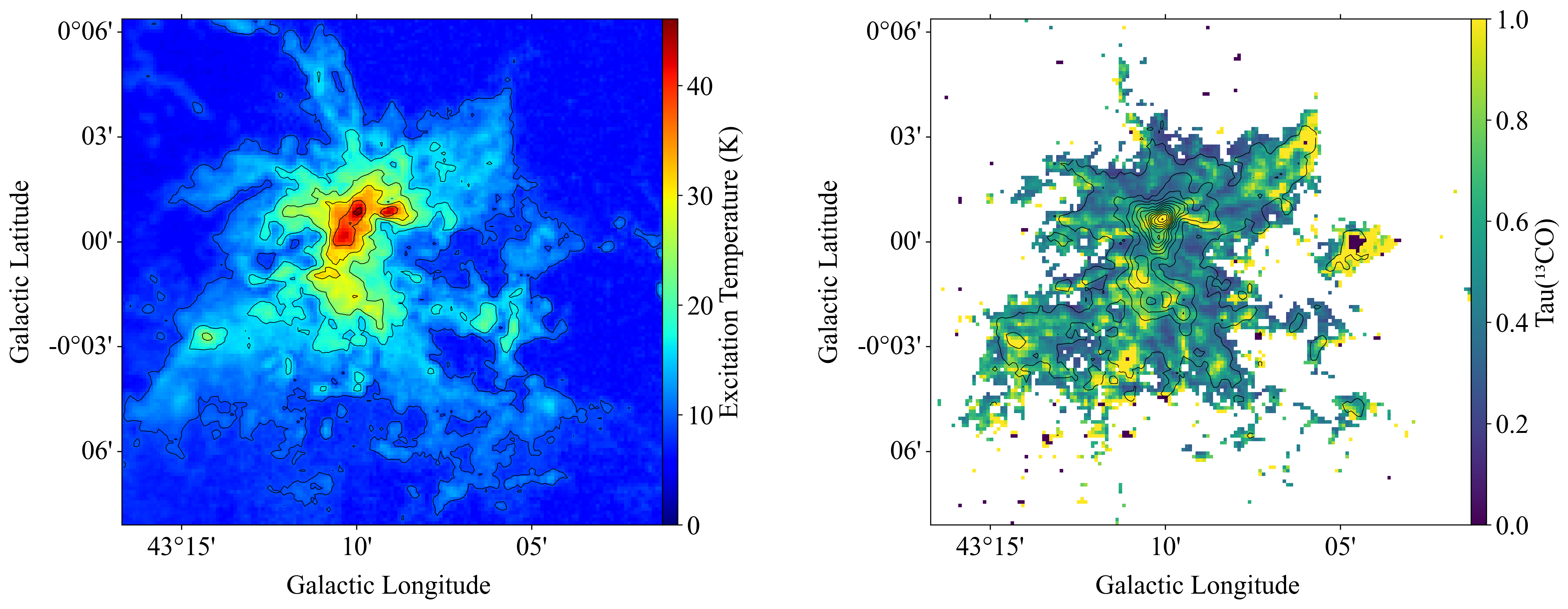}
    \caption{Optical depth in W49A. \textit{Left panel}: Distribution of the excitation temperature. \textit{Right panel}: Optical depth of $^{13}$CO\,(3-2).}
    \label{tao_13CO}
\end{figure*}

Assuming local thermodynamic equilibrium, we estimated the optical depth of $^{13}$CO\,(3-2) using the emission of $^{12}$CO\,(3-2) and $^{13}$CO\,(3-2) with the following formula \citep{paron2018mapping}:
\begin{equation}
    \tau_{^{13}\text{CO}} = -\ln(1-\frac{T_\text{mb}(^{13}\text{CO})}{15.87[\frac{1}{e^{15.87/T_\text{ex}}-1}-0.0028]}).
\end{equation}
The $T_\text{mb}(^{13}\text{CO})$ here is the main-beam brightness temperature, $T_\text{ex}$ is the excitation temperature, assuming $^{12}$CO\,(3-2) emission is optically thick, $T_\text{ex}$ is obtained from
\begin{equation}
    T_\text{ex} = \frac{16.6}{\ln[1+16.6/(T_\text{peak}(^{12}\text{CO})+0.036)]},
\end{equation}
where $T_\text{peak}(^{12}\text{CO})$ is the peak main brightness temperature obtained from the $^{12}$CO\,(3-2) line. Fig.\ref{tao_13CO} shows the distributions of excitation temperature (left panel) and optical depth (right panel) of $^{13}$CO\,(3-2) in W49A. For $^{13}$CO\,(3-2), the optical depth in most regions is less than 1, indicating that analyzing the velocity gradient in W49A using $^{13}$CO\,(3-2) is reliable. We set the upper limit of the color bar range to 1 to better distinguish regions with optical depth higher than 1, they are few and mainly appear as bad pixels in the outskirts of the cloud, which could be an artifact caused by a very low inferred $T_\text{ex}$ or a low signal-to-noise ratio in the $^{13}$CO\,(3-2) data at those positions.
\newpage

\section{Channel maps of $^{13}$CO\,(3-2)}
\label{map}

This part mainly shows the velocity channel map in the W49A area.

\begin{figure*}[hbt]
    \centering
    \includegraphics[width=\textwidth]{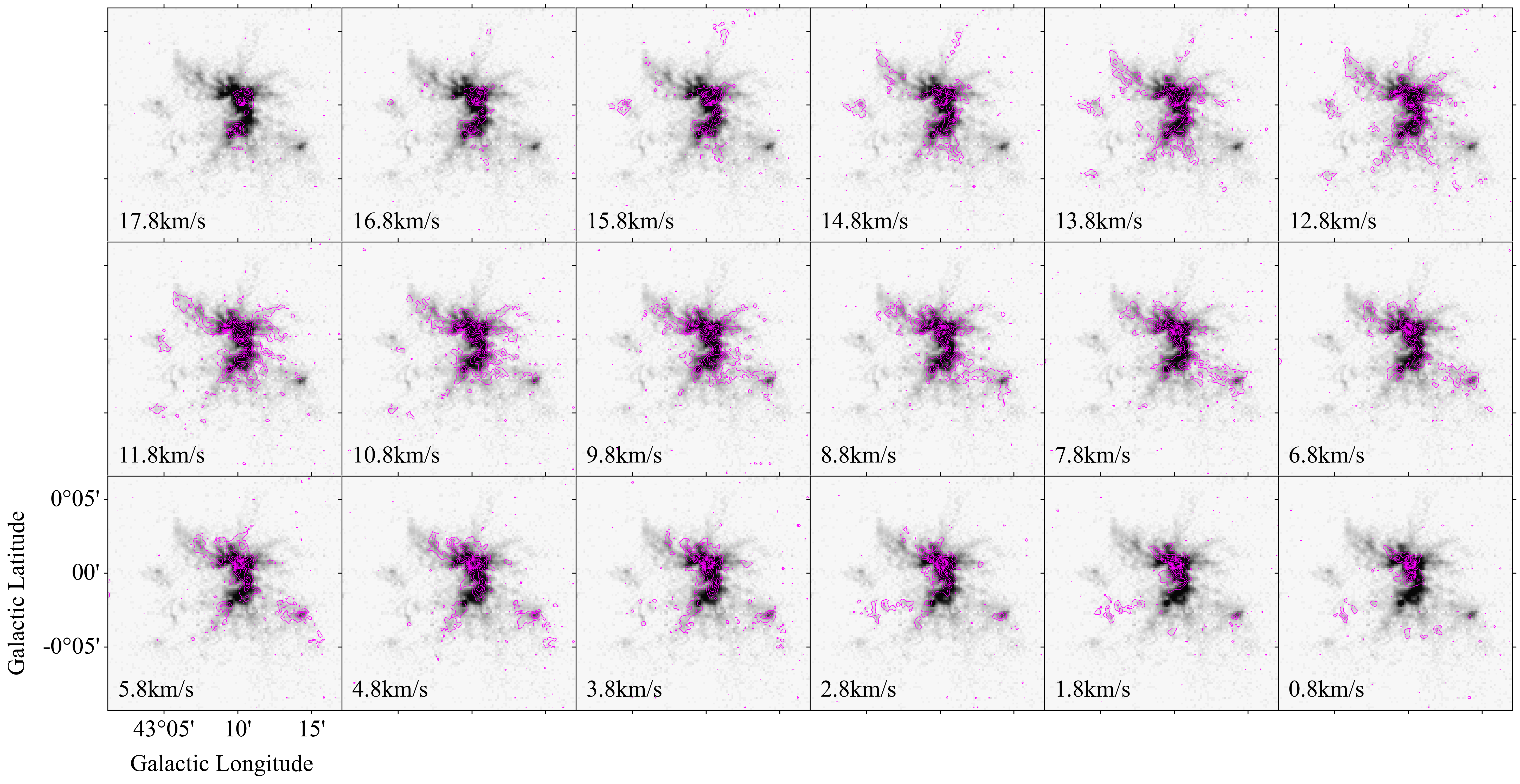}
    \caption{Channel map of $^{13}$CO\,(3-2) in W49A. The color map is the integrated intensity of $^{13}$CO\,(3-2).}
    \label{CO channel map}
\end{figure*}

\newpage
\section{Spectra of some structures in W49A}
\label{dense_clump_spec}

This section mainly presents the spectrum distribution at different positions in W49A, aiming to explore the velocity structure in W49A. It shows the average spectra of $^{13}$CO\,(3-2) at several dense regions in W49A, including the main body of W49A, W49A-N, and the position of strongest $^{13}$CO\,(3-2) emission. It also includes the average CO spectra of dense structures in the B-S and R-S components of W49A-N, the $^{13}$CO\,(3-2) spectra at various positions on filaments in W49A, and the distribution of H$_\alpha$(n=151\,-\,158) RRL at the position of strongest emission in W49A-N.

\begin{figure}[hbt]
    \centering
    \includegraphics[width=1.0\textwidth]{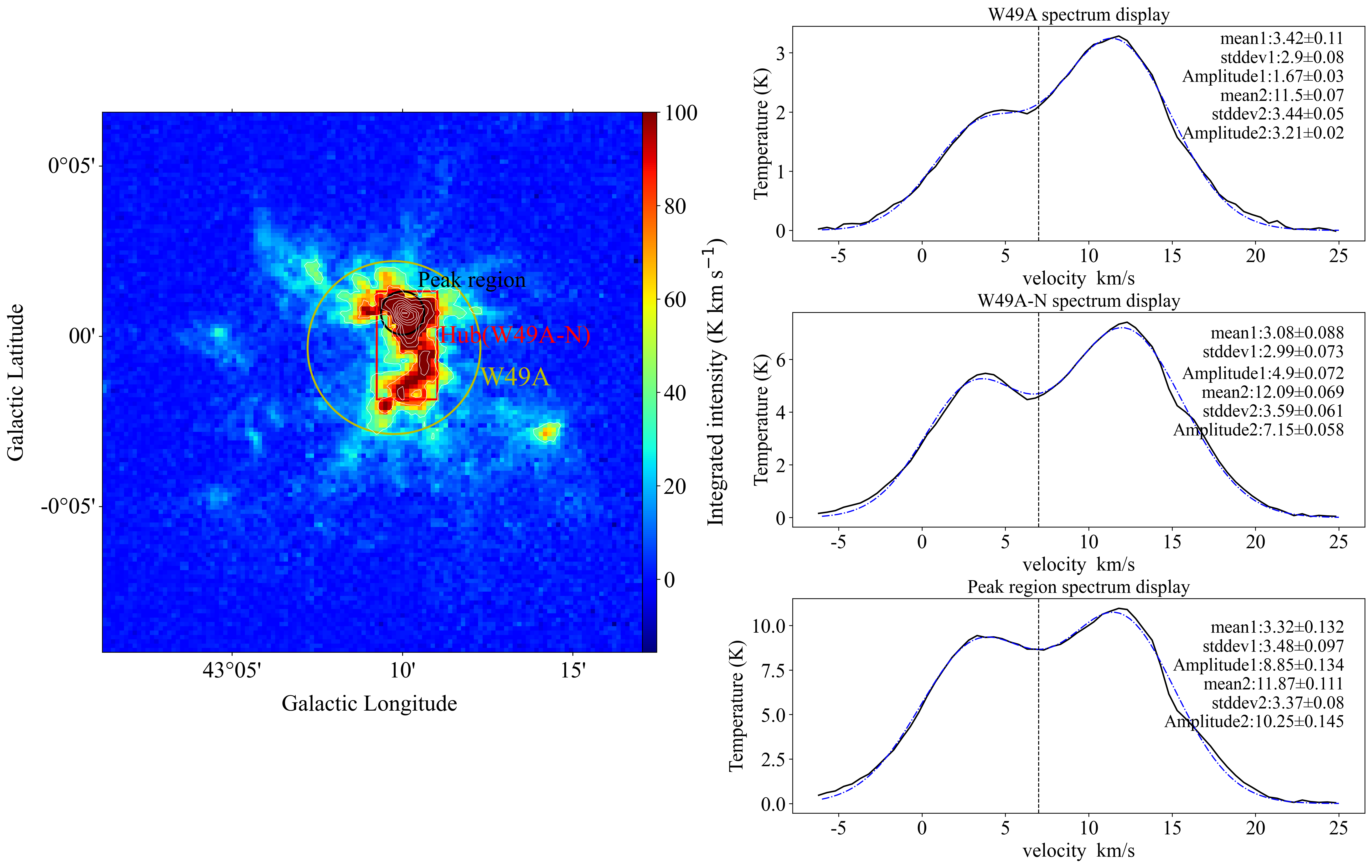}
    \caption{Spectral line analysis of different regions in W49A. \textit{Left panel}: Color map of the integral intensity of $^{13}$CO\,(3-2). Circles and boxes indicate the range of W49A, W49A-N, and the peak region. \textit{Right panel}: Average molecular spectra of $^{13}$CO\,(3-2) for three dense structures. }
    \label{dense spec}
\end{figure}

\begin{figure}[h]
    \centering
    \includegraphics[width=1.0\textwidth]{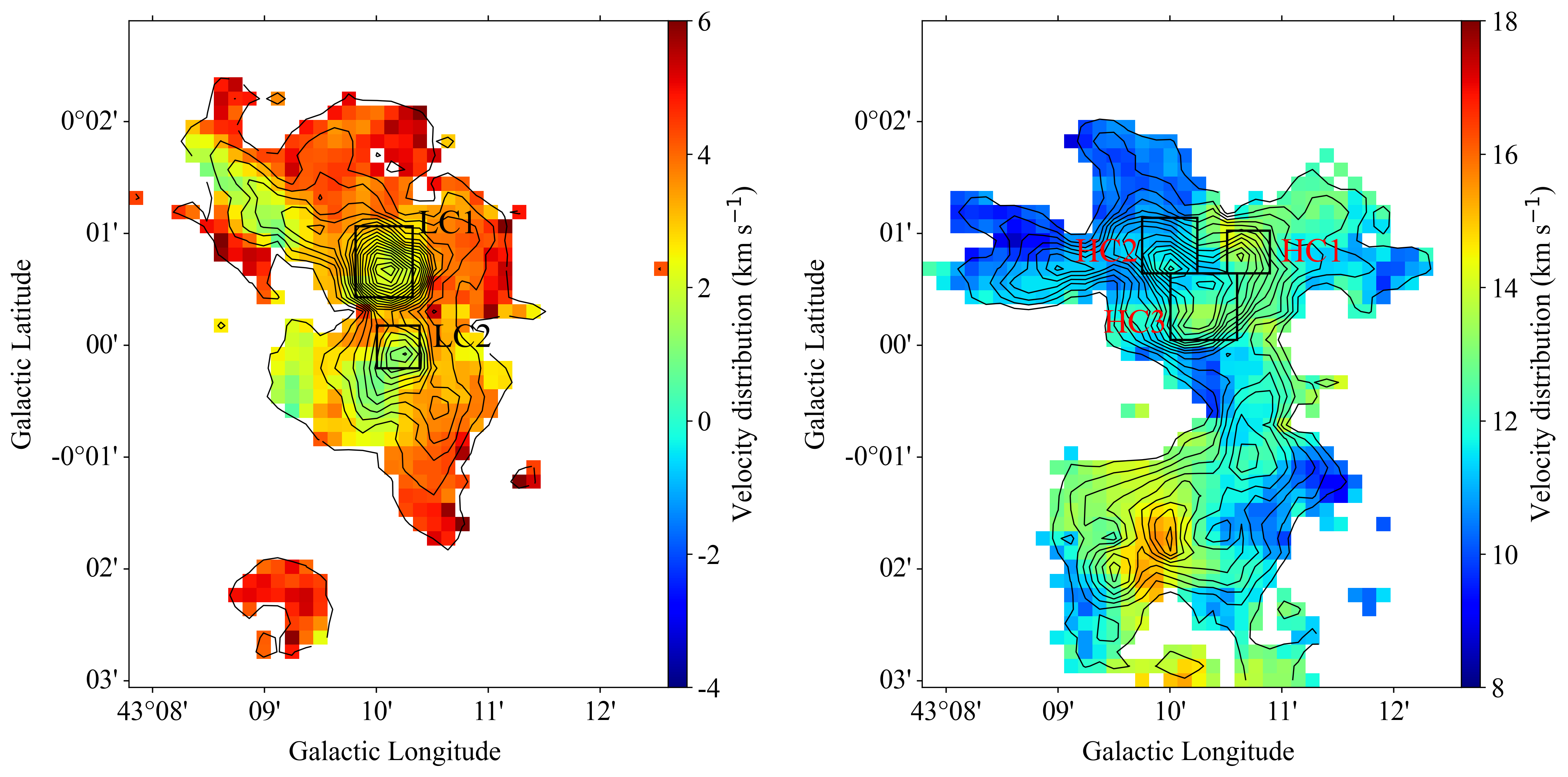}
    \caption{Dense clumps in the hubs of B-S and R-S. \textit{Left panel}: Color map of the $^{13}$CO\,(3-2) velocity distribution of the B-S component. The black outlines are the intensity contours, and the black boxes are the extents of the two dense clumps. 
    \textit{Right panel}: Color map of the $^{13}$CO\,(3-2) velocity distribution of the R-S component. The black outlines are the intensity contours, and the black boxes are the extent of the three dense clumps of the R-S component.}
    \label{low high v}
\end{figure}

\begin{figure}[h!]
    \centering
    \includegraphics[width=1.0\textwidth]{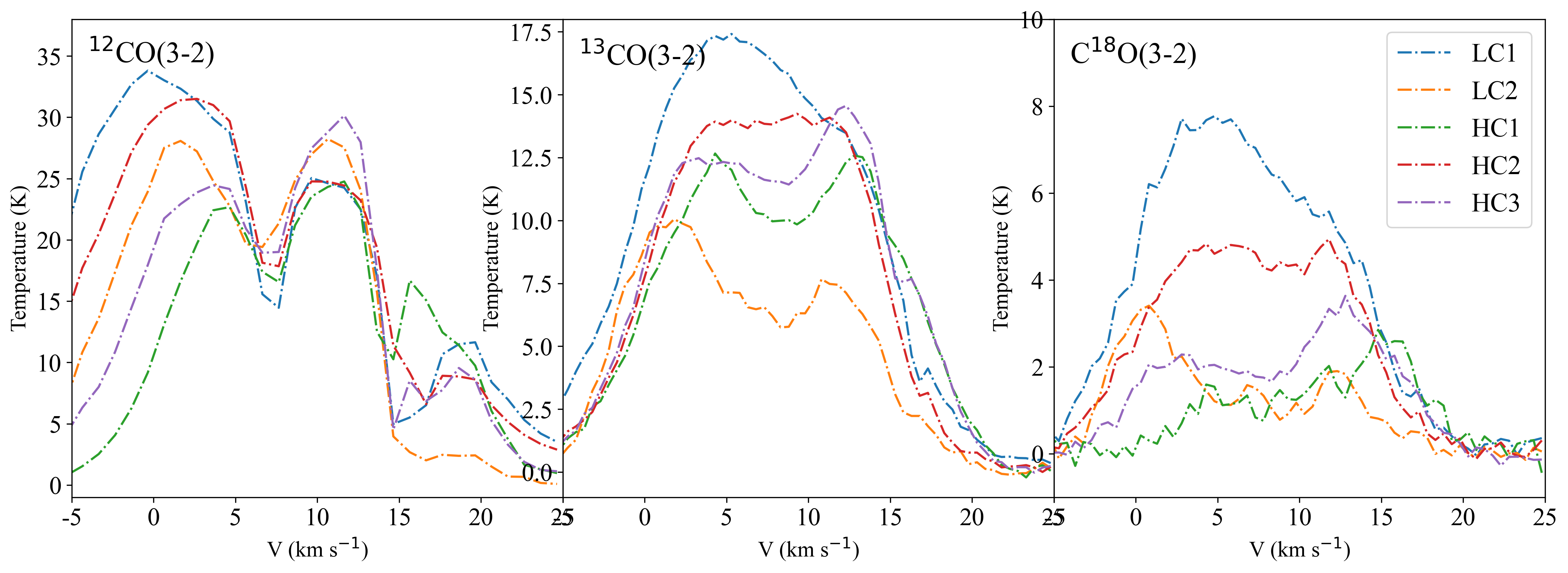}
    \caption{From left to right: Averaged spectra of $^{12}$CO\,(3-2), $^{13}$CO\,(3-2), and C$^{18}$O\,(3-2) corresponding to the positions labeled with clumps in Fig.\ref{low high v}.}
    \label{dense core spec}
\end{figure}

\begin{figure}[h!]
    \centering
    \includegraphics[width=1.0\textwidth]{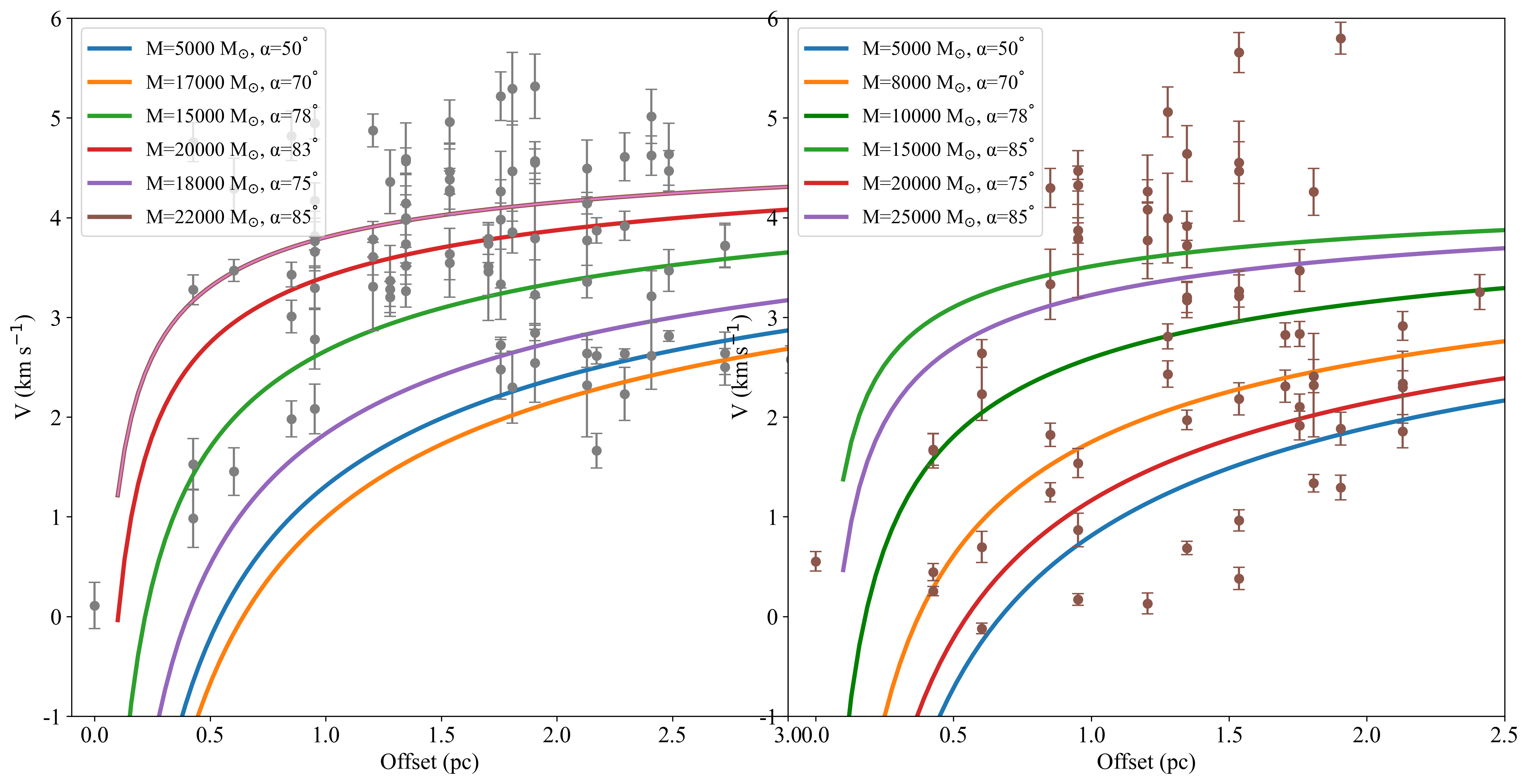}
    \caption{Gravitational collapse of dense clumps in the hub. According to Eqs.(\ref{eq.8}) and (\ref{eq.9}), to test whether the two dense clumps (LC\,1 and LC\,2) in the B-S HFS are in a state of gravitational collapse, we examined their velocity distributions with radius. \textit{Left panel}: Velocity distribution with radius in LC\,1 ($l_0$ = 43.17$^{\circ}$, $b_0$ = 0.00$^{\circ}$), with a fitted curve yielding a systemic velocity of 5.0\,km\,s$\rm ^{-1}$. Based on the integral intensity weighted by the B-S component, we estimated the clump mass to be approximately 17997\,M$_{\odot}$, with the best-fit result at 15000\,M$_{\odot}$, and an inclination angle of 78$^{\circ}$ for the plane of the sky. \textit{Right panel}: Velocity distribution with radius in LC\,2 ($l_0$ = 43.17$^{\circ}$, $b_0$ = -0.00$^{\circ}$), with a fitted curve yielding a systemic velocity of 4.5\,km\,s$\rm ^{-1}$. Based on the integral intensity weighted by the B-S component, we estimate the clump mass to be approximately 9098\,M$_{\odot}$, with the best-fit result at 10000\,M$_{\odot}$, and an inclination angle of 78$^{\circ}$ concerning the plane of the sky.}
    \label{lc1_lc2}
\end{figure}

\begin{figure}[h!]
    \centering
    \includegraphics[width=1.0\textwidth]{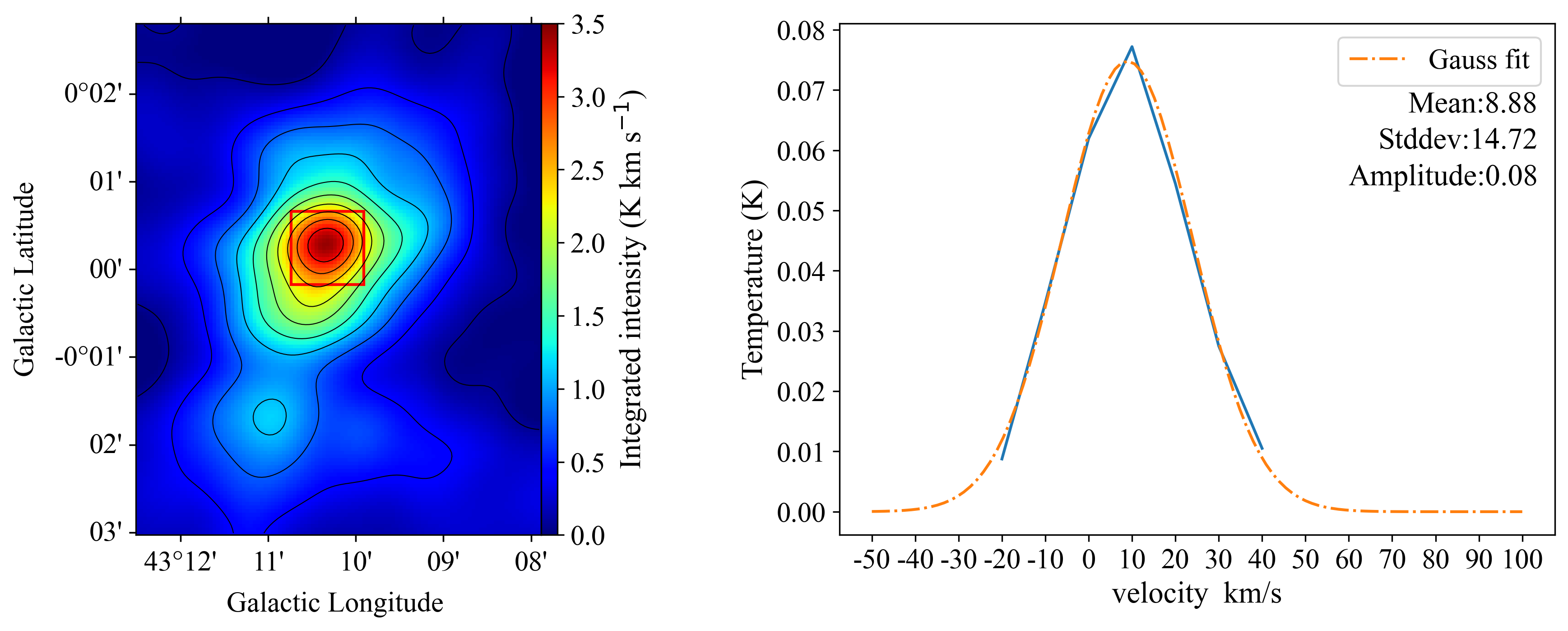}
    \caption{RRL at the position of the strongest emission in W49A-N. \textit{Left panel}: Distribution of H$_\alpha$(n=151\,-\,158) emission. \textit{Right panel}: Average spectrum of dense regions, representing the velocity distribution of ionized gas.}
    \label{rrl}
\end{figure}

\begin{figure}[h!]
    \centering \includegraphics[width=1.0\textwidth]{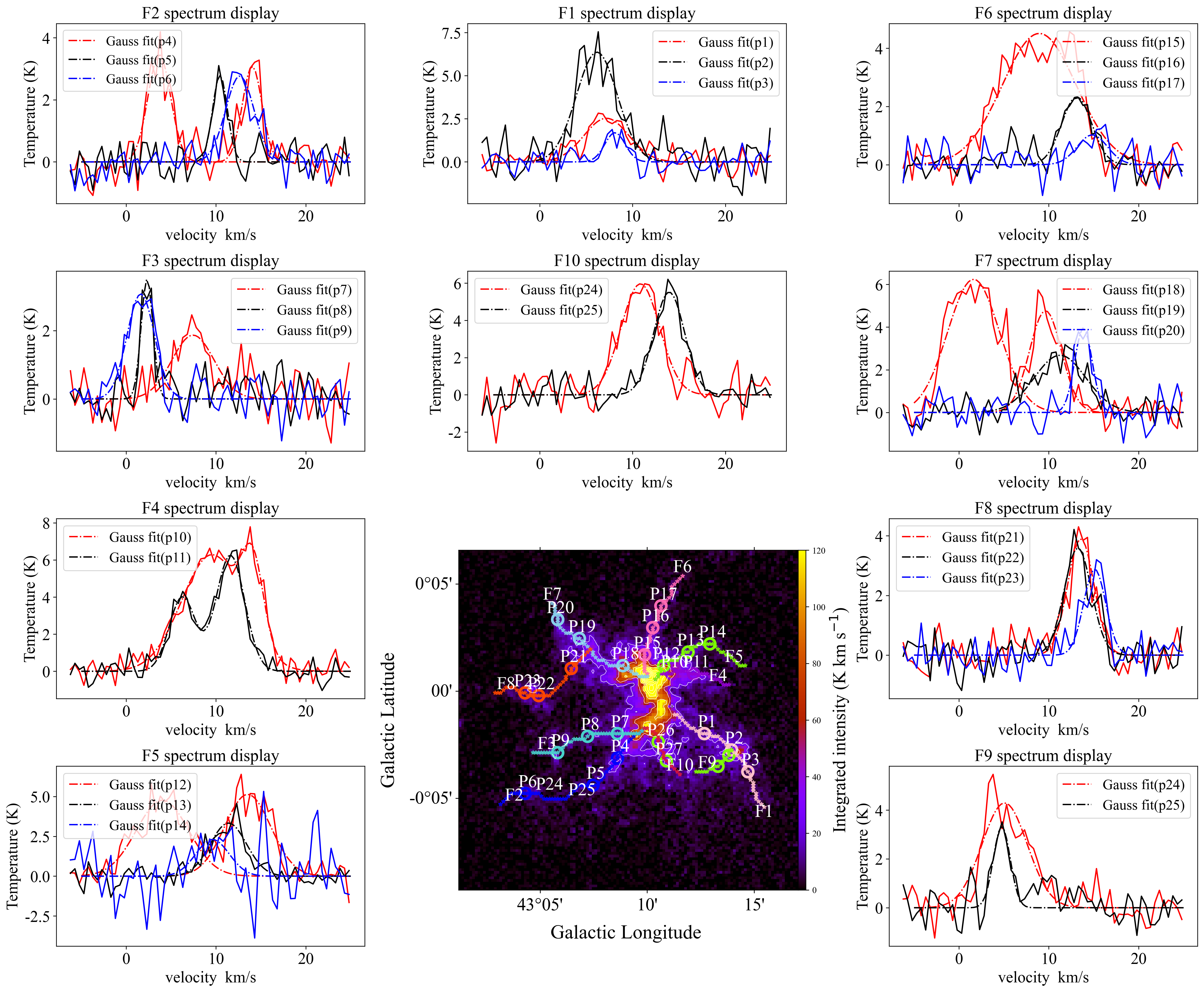}
    \caption{Spectra of $^{13}$CO\,(3-2) along ten different filaments at the positions marked in the central figure.}
   \label{filament_spec}
\end{figure}

\newpage
\onecolumn
\section{Position-velocity diagram along the filaments}
\label{P-V diagram}
Here we present the P-V diagram of filaments in W49A. We performed Gaussian fitting to the averaged $^{13}$CO\,(3-2) spectra within a 9-pixel region around each pixel along the skeleton of the filaments. This process allows us to obtain the peak velocity and the fitting error at each position along the filament. It is worth noting that there is a noticeable double-peak structure near the W49A-N region. For these positions, we conducted Gaussian double-peak fitting (see Eq.(\ref{eq:bimodal}) ) to obtain velocities for both the B-S and R-S filaments: 
\begin{equation}
     f(x) = A_{1} e^{-\frac{(x - \mu_{1})^2}{2\sigma_{1}^2}} + A_{2} e^{-\frac{(x - \mu_{2})^2}{2\sigma_{2}^2}}.
    \label{eq:bimodal}
\end{equation}
However, the two velocity components are mixed in certain areas within the hub without a clear double-peak structure. In such cases, our approach was to conduct single-peak Gaussian fitting (see Eq.(\ref{eq:gauss}) ) to the mixed components:
\begin{equation}
     f(x) = A e^{-\frac{(x - \mu)^2}{2\sigma^2}}.
    \label{eq:gauss}
\end{equation}

Meanwhile, we marked the positions corresponding to dense structures on each P-V diagram and calculated the velocity gradient along the filament through linear fitting.

\begin{figure*}[hbt]
    \centering
    \includegraphics[width=\textwidth]{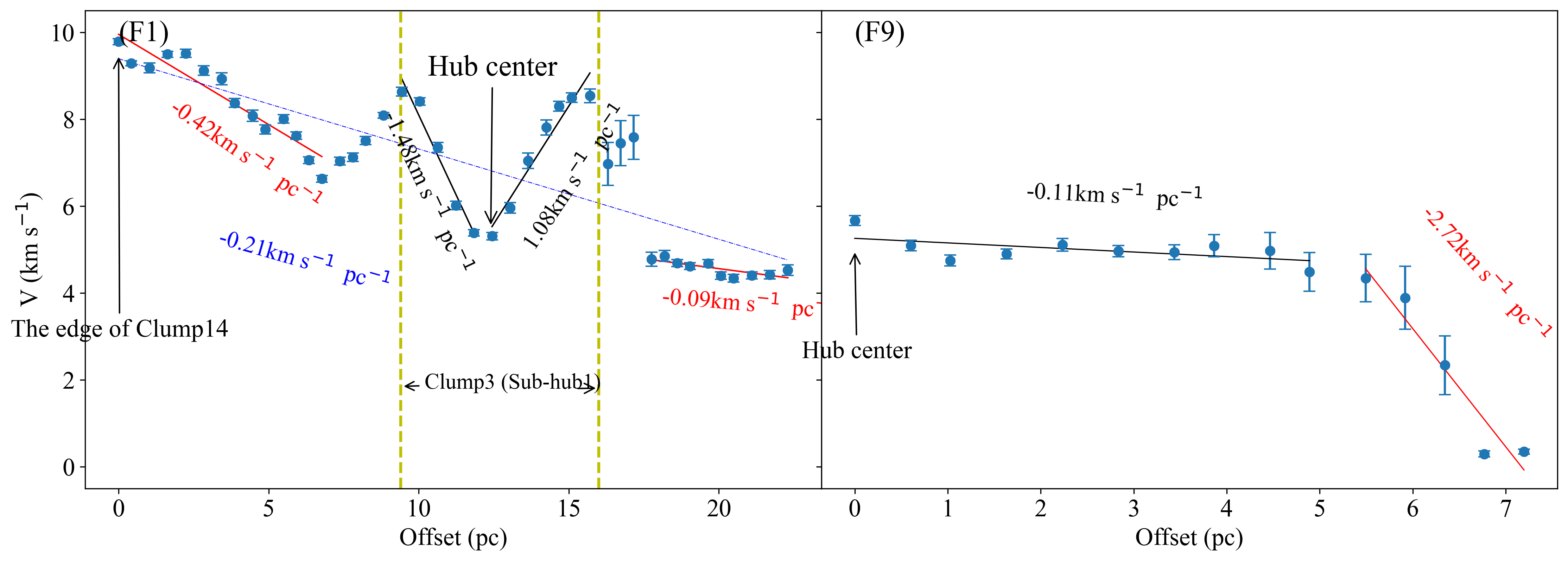}
    \caption{P-V diagrams and velocity gradients of two filaments, F\,1 (left panel) and F\,9 (right panel).}
    \label{f1_f9_pv}
\end{figure*}

\begin{figure*}[hbt]
    \centering
    \includegraphics[width=\textwidth]{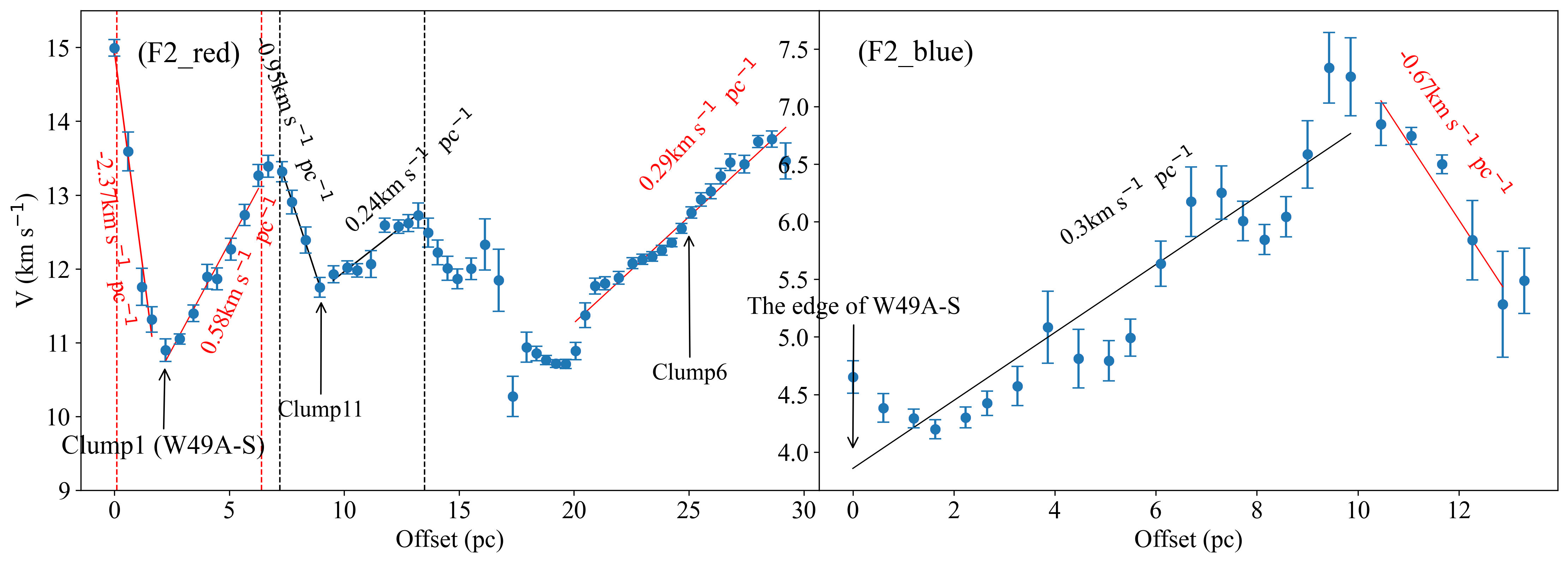}
    \caption{P-V diagrams and velocity gradients of filament F\,2, R-S (left panel) and B-S (right panel).}
    \label{f2_pv}
\end{figure*}

\begin{figure*}[hbt]
    \centering
    \includegraphics[width=\textwidth]{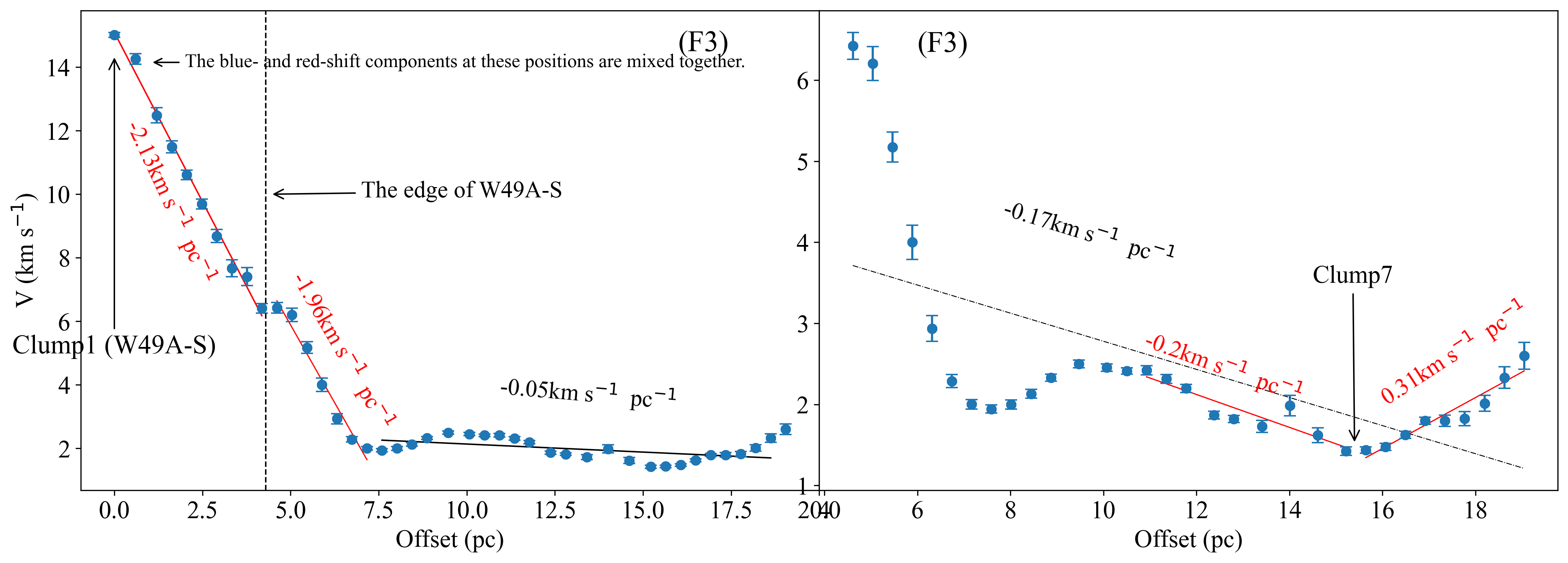}
    \caption{P-V diagrams and velocity gradients of filament F\,3.}
    \label{f3_pv}
\end{figure*}

\begin{figure*}[hbt]
    \centering
    \includegraphics[width=\textwidth]{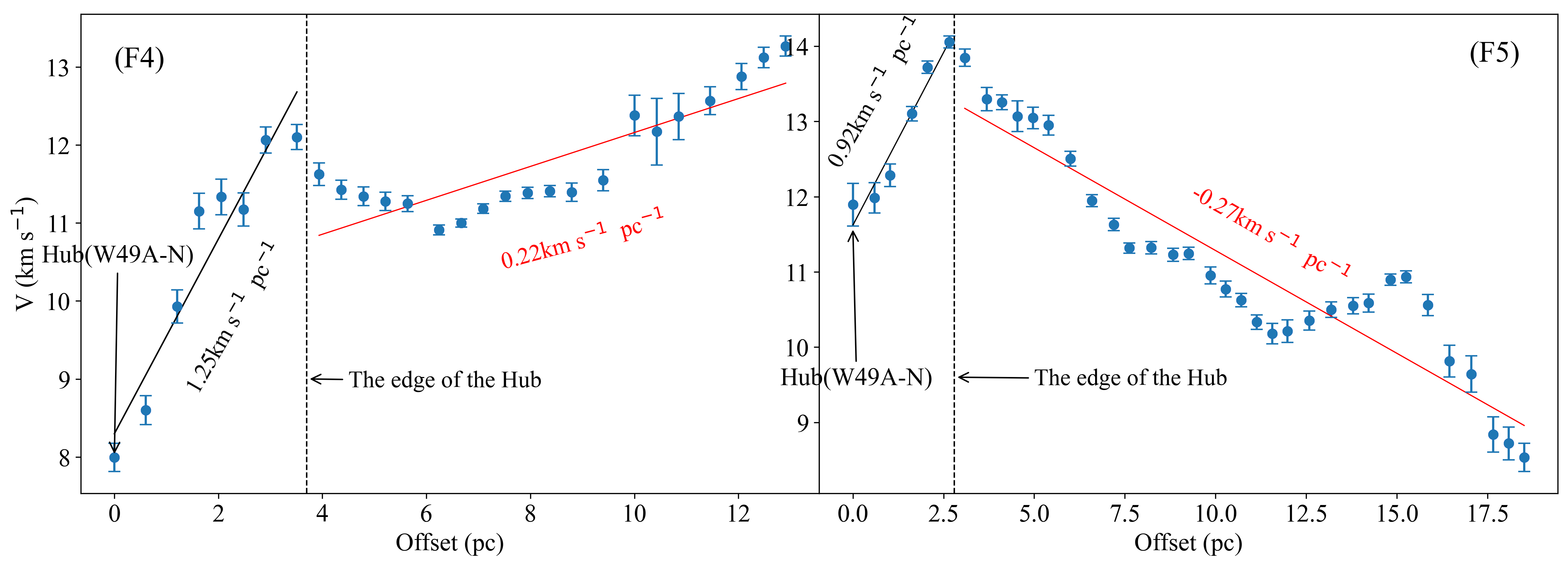}
    \caption{P-V diagrams and velocity gradients of two filaments, F\,4 (left panel) and F\,5 (right panel).}
    \label{f4_f5_pv}
\end{figure*}

\begin{figure*}[hbt]
    \centering
    \includegraphics[width=\textwidth]{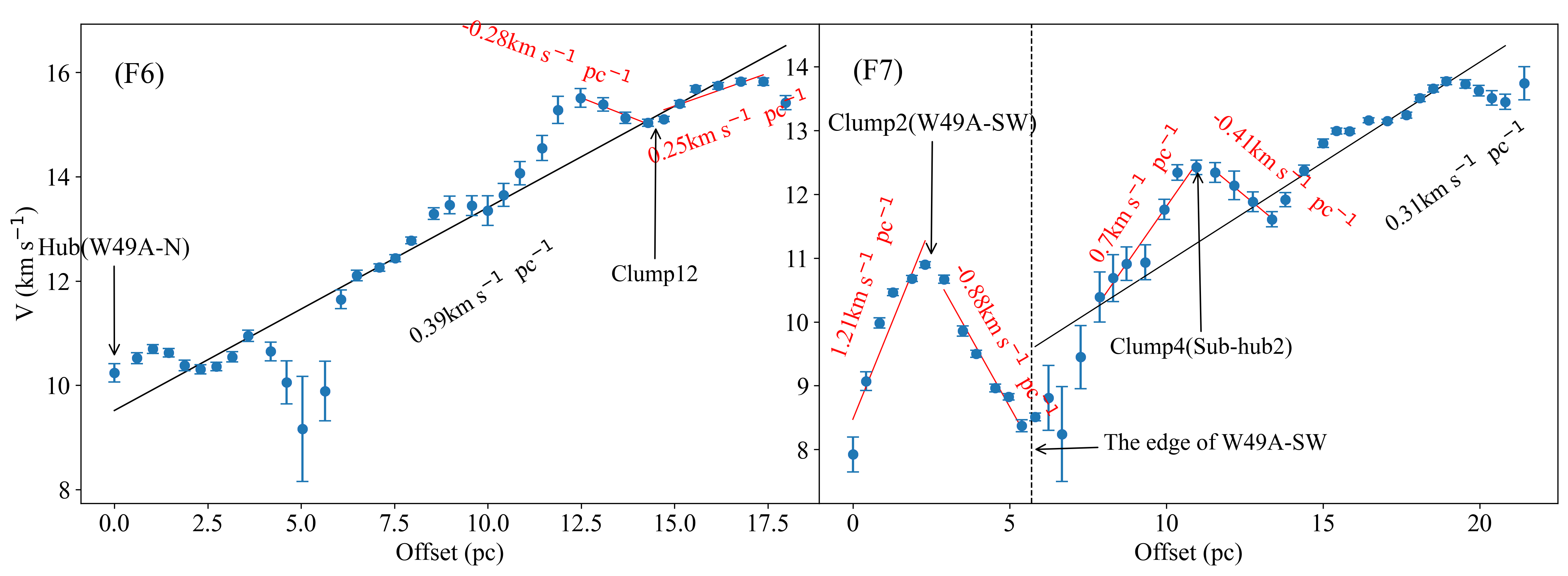}
    \caption{P-V diagrams and velocity gradients of two filaments, F\,6 (left panel) and F\,7 (right panel).}
    \label{f6_f7_pv}
\end{figure*}

\begin{figure*}[hbt]
    \centering
    \includegraphics[width=\textwidth]{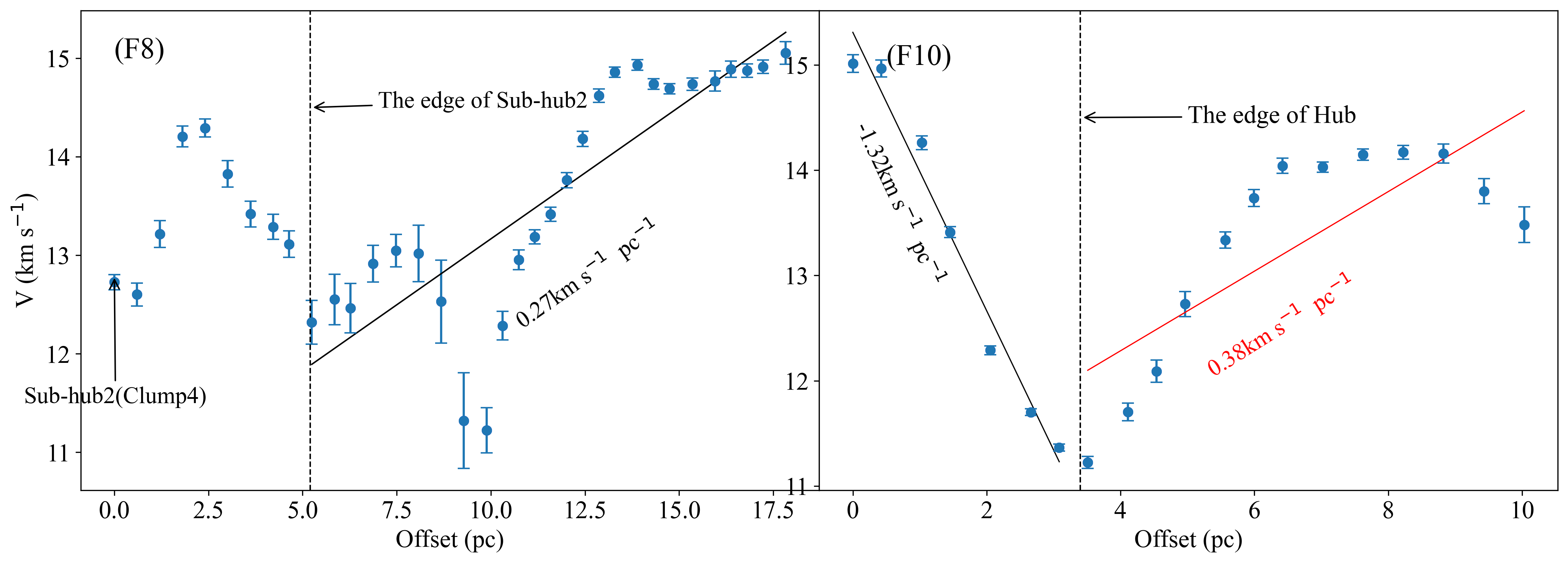}
    \caption{P-V diagrams and velocity gradients of two filaments, F\,8 (left panel) and F\,10 (right panel).}
    \label{f8_f10_pv}
\end{figure*}

\end{appendix}

\end{document}